\documentclass[11pt]{article}
\usepackage{graphicx} % Required for inserting images
\usepackage{amsmath}
\usepackage{amssymb}
\usepackage{xcolor}
\usepackage{float} 
\usepackage{overpic}
\usepackage{subcaption}
\usepackage{csquotes}
\usepackage{setspace}
\usepackage{todonotes}
\usepackage{ulem}

\usepackage{placeins}

\usepackage[a4paper, margin=1in]{geometry} 
\numberwithin{equation}{section}
\usepackage[hidelinks]{hyperref}
\usepackage[sort&compress]{cleveref}
\crefname{table}{Table}{Tables}
\crefname{appendix}{Appendix}{Appendices}
\crefname{figure}{Figure}{Figures}
\crefname{section}{Section}{Sections}  
\crefname{chapter}{Chapter}{Chapters}  
\crefname{equation}{}{}
\creflabelformat{equation}{[#2#1#3]}
\usepackage{tcolorbox}
\usepackage[
  backend=biber,
  style=numeric-comp,
  sorting=none,
  giveninits=true,
  date=year,
  url=false,
  isbn=false,
  maxnames=5,
  minnames=4,
  maxcitenames=2,
  mincitenames=1,
  autocite=superscript
]{biblatex}

\DeclareBibliographyDriver{article}{%
  \usebibmacro{bibindex}%
  \usebibmacro{begentry}%
  \usebibmacro{author/editor+others}%
  \setunit{\labelnamepunct}\newblock
  \usebibmacro{title}%
  \newunit\newblock
  \usebibmacro{journal}%
 \setunit{\addcomma\space}%
  \usebibmacro{volume+number+eid}%
  \setunit{\addcomma\space}%
  \printfield{pages}%
  \setunit{\space}%
  \printfield{year}%
  \newunit\newblock
  \usebibmacro{finentry}%
}

\DeclareBibliographyDriver{book}{%
  \usebibmacro{bibindex}%
  \usebibmacro{begentry}%
  \usebibmacro{author/editor+others}%
  \setunit{\labelnamepunct}\newblock
  \usebibmacro{title}%
  \newunit\newblock
  \printlist{publisher}%
  \setunit{\addcomma\space}%
  \printlist{location}%
  \setunit{\space}%
  \printfield{year}%
  \newunit\newblock
  \usebibmacro{finentry}%
}

\DeclareFieldFormat{title}{\uline{#1\isdot}}

\renewbibmacro*{title}{%
  \printfield[title]{title}%
  \newunit
}
\AtEveryBibitem{%  
  \clearfield{issn}   
  \clearfield{isbn}  
  \clearfield{eid}%
  \clearfield{url}    
  \clearfield{urldate} % date visited
  \clearfield{language} 
  \clearfield{langid} 
  \clearfield{number}%
}
\DeclareFieldFormat[article]{pages}{#1}

\DeclareFieldFormat[article]{volume}{\mkbibbold{#1}}
\DeclareFieldFormat[article]{number}{\mkbibparens{#1}}
\DeclareFieldFormat{year}{\mkbibparens{#1}}
\renewcommand*{\labelnamepunct}{\addcomma\space}

\DeclareFieldFormat[article]{pages}{\mkfirstpage{#1}}
\renewbibmacro*{volume+number+eid}{%
  \printfield{volume}%
  \iffieldundef{number}
    {}
    {\setunit{\space}\printfield{number}}%
}
\renewbibmacro*{in:}{%
  \ifentrytype{article}
    {}
    {\printtext{\bibstring{in}\intitlepunct}}%
}
\DeclareBibliographyDriver{software}{%
  \usebibmacro{bibindex}%
  \usebibmacro{begentry}%
  \usebibmacro{author/editor+others}%
  \setunit{\labelnamepunct}\newblock
  \usebibmacro{title}%
  \newunit\newblock
  \printfield{version}%
  \newunit\newblock
  \printfield{year}%
  \newunit\newblock
  \iftoggle{bbx:doi}
    {\printfield{doi}}
    {}%
  \newunit\newblock
  \iftoggle{bbx:url}
    {\printfield{url}}
    {}%
  \usebibmacro{finentry}%
}

\DeclareBibliographyDriver{incollection}{%
  \usebibmacro{bibindex}%
  \usebibmacro{begentry}%
  \usebibmacro{author}%
  \setunit{\addcomma\space}%
  \printtext{in\space}%
  \printfield{booktitle}
  \setunit{\addcomma\space}%
  \printnames{editor}%
  \setunit{\addcomma\space}%
  \ifnumgreater{\value{editor}}{1}
    {\printtext{Editors}}
    {\printtext{Editor}}%
  \setunit{\addcomma\space}%
  \printfield{pages}
  \setunit{\addcomma\space}%
  \printlist{publisher}%
  \setunit{\addcomma\space}%
  \printlist{location}%
  \setunit{\space}%
  \printfield{year}%
  \addperiod
  \usebibmacro{finentry}%
}

\let\citep\parencite
\DeclareCiteCommand{\citet}
  {}
  {%
    \printnames{labelname}%
    \autocite{\thefield{entrykey}}%
  }
  {\multicitedelim}
  {}

\usepackage{xstring}
\usepackage{refcount}   % provides \getrefnumber{<label>}
\usepackage{hyperref}   % optional, only if you want clickable citations
\makeatletter
\AddToHook{cmd/appendix/before}{\def\cref@section@alias{appendix}\def\cref@subsection@alias{appendix}}
\makeatother

\newcommand{\refstriplast}[2][2]{%
  \begingroup
    \edef\refstriptemp{\getrefnumber{#2}}% get raw stored ref text
    \StrLen{\refstriptemp}[\refstriplen]%
    \StrLeft{\refstriptemp}{\number\numexpr\refstriplen-#1\relax}%
  \endgroup
}

\newcommand{\hrefpart}[2]{% #1 = label, #2 = letter (a or b)
  \hyperref[#1]{(\refstriplast{#1}#2)}%
}

\newcommand{\R}{\hspace{-0.8 mm}:\hspace{-0.8 mm}}

\newcommand{\dint}[4]{\int_{#1}^{#2}\! #3 \, \mathrm{d}#4}
\newcommand{\eps}{\varepsilon}
\newcommand{\ee}[1]{\mathrm{e}^{#1}}
\newcommand{\dd}[2]{\frac{\mathrm{d} {#1}}{\mathrm{d} {#2}}}
\newcommand{\pOG}{p}
\newcommand{\nOG}{n}
\newcommand{\potOG}{\phi}
\newcommand{\p}{p_0}
\newcommand{\n}{n_0}
\newcommand{\C}{c_0}
\newcommand{\rr}{r}
\newcommand{\mL}{u_L}
\newcommand{\mR}{u_R}

\newcommand{\Kc}{\kappa}
\newcommand{\pot}{\phi_0}
\newcommand{\Ip}{I_p}
\newcommand{\II}{\mathcal{I}}

\newcommand{\NN}{\gamma}
\newcommand{\In}{I_n}
\newcommand{\critpot}{\bar{\phi}}

\newcommand{\B}{\mathcal{J}}
\newcommand{\tl}{\alpha_L} % new definition for asym case constant 
\newcommand{\tr}{\alpha_R}

\newcommand{\pL}{P}
\newcommand{\nL}{N}
\newcommand{\potL}{\Phi}
\newcommand{\pR}{\mathcal{P}}
\newcommand{\nR}{\mathcal{N}}
\newcommand{\potR}{\varphi}
\newcommand{\xL}{\xi}
\newcommand{\xR}{\eta}

\usepackage{chngcntr}        % safe even if already loaded
\counterwithout{equation}{section} % stop tying eq counter to sections
\renewcommand{\theequation}{\arabic{equation}} % display as (1), (2), ...
\allowdisplaybreaks

\makeatletter
\renewcommand{\tagform@}[1]{[#1]}
\makeatother

\begin{document}
\DeclareQuotePunctuation{.,}

%\input{JESAuthorGuide}
%\newpage
\title{Modelling Intermediate-Current Transitions in Asymmetric-Valence Binary Electrolytes}
\author{
Georgina C. Ryan\textsuperscript{z},
Mohit P. Dalwadi,
Ian M. Griffiths
\\[1ex]
Mathematical Institute, University of Oxford, Oxford OX2 6GG, UK
}
\footnotetext{Email: georgina.ryan@maths.ox.ac.uk}
\date{\vspace{-5ex}}
\maketitle

\begin{abstract}
Asymmetric valences in a binary electrolyte can significantly affect the performance of systems such as reverse electrodialysis cells, batteries, and supercapacitors. To generate a theoretical understanding of this effect, we consider a steady one-dimensional Poisson--Nernst--Planck model of an electrolytic cell with imposed constant ionic fluxes, focusing on varying ion valences in a general asymmetric binary electrolyte. Numerical simulations reveal a smooth transition between the qualitatively distinct near-equilibrium and strongly non-equilibrium steady-state regimes.  These regimes are distinguished by a valence-dependent transition point at an intermediate current where the classical Debye-scale boundary layer vanishes. We characterise this transition using asymptotic analysis, recovering the Gouy--Chapman and limiting-current results in the appropriate limits, and determining the correct transition results when neither is appropriate. We provide implicit solutions for the potential and ion concentrations of general asymmetric binary electrolytes and, notably, we provide explicit analytic expressions for the asymptotic composite solutions for $2z\R z$, $z\R 2z$, and $z\R z$ electrolytes. We show how the results can be presented in a collapsed phase diagram that can be used to predict qualitative intermediate-current steady-state behaviour in terms of ion valences and fluxes.
\end{abstract}

\section*{Introduction}

The presence of multivalent ions can have a significant impact on the performance of electrochemical systems. For example, in reverse electrodialysis systems, where electricity is generated by mixing concentrated and dilute sodium chloride solutions, adding divalent ions to the electrolyte mixture reduces the open circuit voltage \autocite{beshaDesignMonovalentIon2019,wuMitigatingInfluenceMultivalent2024,postInfluenceMultivalentIons2009}. Moreover, these reverse electrodialysis systems are designed to generate energy from mixing natural water sources, which naturally contain multivalent ions \autocite{postInfluenceMultivalentIons2009}. In battery modelling, multivalent battery systems are a growing area of interest due to the potential of higher volumetric energy density than traditional Li-ion batteries \autocite{guduruBriefReviewMultivalent2016,shahReviewRecentAdvances2025}. Additionally, the energy storage in supercapacitors is determined by the electrical double layer (EDL) at the electrolyte/electrode interface, and the properties of the EDL in a supercapacitor vary significantly with the valency of ions in the electrolyte \autocite{mendheReviewElectrolytesSupercapacitor2023,messiasAssessingImpactValence2022, guptaElectricalDoubleLayers2018}. \\

The impact of valence asymmetry on the fundamental properties of a binary electrolyte is a growing area of research. By studying modified Poisson--Boltzmann equations, Gupta et.~al\autocite{guptaElectricalDoubleLayers2018,guptaDiffusiophoreticDiffusioosmoticVelocities2019} found that asymmetric valences significantly impacted steric effects, dielectric decrement, ion--ion correlations, and relative ion motion in an electrolyte (diffusiophoresis and diffusioosmosis). From molecular dynamics simulations, Messias et.~al\autocite{messiasAssessingImpactValence2022,messiasSaltinwaterWaterinsaltElectrolytes2022} identified that asymmetric valence impacts many fundamental electrolyte properties, including EDL capacitance, cohesive energy, viscosity, and diffusion coefficients. Additionally, the coupling of diffusivity and valence asymmetry for the time-dependent Poisson--Nernst--Planck (PNP) equations with blocking-electrode conditions (i.e. not allowing a Faradaic current\autocite{bazantDiffusechargeDynamicsElectrochemical2004}) has been considered in flat-plate\autocite{baluElectrochemicalImpedanceSpectrum2022} and pore\autocite{henriqueImpactAsymmetriesValences2022a} geometries, with notable effects including producing oscillating diffusion layers and changing EDL charging timescales respectively. \\

Despite the noted effects and growing importance of valence asymmetry in electrochemical systems, the majority of mathematical modelling of ion transport in dilute solutions assumes a symmetric \mbox{$z\R z$} electrolyte. This assumption creates symmetries in the governing PNP equations that significantly simplify the mathematics \autocite{bazantCurrentVoltageRelationsElectrochemical2005,newmanElectrochemicalSystems2021}. 
Many foundational papers that have developed the field of asymptotic modelling of the PNP equations use the symmetric electrolyte simplification \autocite{biesheuvelImposedCurrentsGalvanic2009,bazantCurrentVoltageRelationsElectrochemical2005,bazantDiffusechargeDynamicsElectrochemical2004,bonnefontAnalysisDiffuselayerEffects2001}.  We note that solving the general \mbox{$z\R z$} and the \mbox{$1\R1$} electrolyte cases are equivalent under appropriate scalings in the non-dimensionalisation, so valence can be effectively scaled out of the mathematical problem in the symmetric case\autocite{bazantCurrentVoltageRelationsElectrochemical2005}. A simplification of this nature is not possible for asymmetric electrolytes; the valences must play some role in the dimensionless PNP equations. \\

The behaviour of an electrolytic cell can be divided into two distinct regimes: (thermodynamic) equilibrium and non-equilibrium \autocite{singerSingularPerturbationAnalysis2008}. Equilibrium refers to the regime where each ionic species has no net flux at the electrodes and hence there is no Faradaic current\autocite{newmanElectrochemicalSystems2021}. Non-equilibrium refers to systems with a Faradaic current, so there is a nonzero ionic flux at the electrodes and ion transport throughout the bulk. Non-equilibrium systems can be either transient (i.e. described by the time-dependent Poisson--Nernst--Planck equations) or steady. We use the term `strongly non-equilibrium' to refer to systems that are approaching the limiting current, at which an electroactive ion is depleted to zero concentration at the electrode where it is consumed \autocite{newmanElectrochemicalSystems2021}.\\

Asymmetric valence electrolytes have been well-studied in the equilibrium case.  In equilibrium, Gouy--Chapman theory describes the potential in the diffuse-layer of the EDL near an electrode as a solution to the Poisson--Boltzmann equation, a simplification of the PNP equations\autocite{gouyConstitutionChargeElectrique1910,chapmanLIContributionTheory1913}.  In the diffuse layer, the concentrations of both ion species are determined by the potential via a Boltzmann relation that depends on the ion valence\autocite{bardElectrochemicalMethodsFundamentals2022}. There are only a few examples of analytic solutions to the Poisson--Boltzmann equation for asymmetric valence electrolytes in terms of elementary functions, namely the classic solutions for \mbox{$z\R 2z$} and \mbox{$2z\R z$} electrolytes at a single blocking electrode  \autocite{gouyConstitutionChargeElectrique1910,andriettiExactSolutionUnidimensional1976}. Analytic solutions for parallel-plate geometry have also been identified for symmetric and \mbox{$2z\R z$} electrolytes in terms of Jacobi and Weierstrass elliptic functions respectively \autocite{zhangExactSolutionNonlinear2018,xingPoissonBoltzmannTheoryTwo2011}.\\

Two studies relevant to our model have considered valence asymmetry in non-equilibrium systems. Firstly, \citet{jarveyIonTransportElectrochemical2022} model the time-evolution of a general-valence PNP system, focusing on the coupling of EDL charging and redox reactions. The authors use asymptotic analysis to derive a set of differential equations that are much simpler to solve numerically compared to simulating the full PNP system. Since Jarvey et al. consider the time-evolution problem, their composite matching solutions for the concentrations and potential are given implicitly in the form of PDEs (bulk region) and ODEs and algebraic constraints (diffuse layers) alongside boundary and initial conditions.
The model explicitly includes a linear potential drop to model the Stern layer of the EDL and models redox reactions by setting the flux, which may be a function of the concentration and potential, at the boundary of the Stern and diffuse layers (i.e. the `outer-sphere approximation' \autocite{newmanElectrochemicalSystems2021}). 
\\

Secondly,
\citet{wangSingularPerturbationSolutions2014} present a singular perturbation analysis of the PNP equations to find the steady-state electric potential of a general number of ion species and general valences. Their model uses Dirichlet concentration boundary conditions and unknown fluxes, a common choice for modelling ion channels in biological cells \autocite{kellerModelFrameworkIon2025,zhengSecondorderPoissonNernst2011a,singerSingularPerturbationAnalysis2008}. The solutions are presented in terms of implicit integrals and inverse functions, and are shown to agree with the classic \mbox{$1\R 1$} asymptotic results of \citet{chenQualitativePropertiesSteadyState1997} for Dirichlet concentration boundary conditions.\\

In this paper, we consider the steady state of a one-dimensional Poisson--Nernst--Planck model of an electrolytic cell. In contrast to \citet{wangSingularPerturbationSolutions2014}, we consider imposed constant ionic flux boundary conditions, focusing on varying ion valences in a general asymmetric binary electrolyte. 
This choice is more appropriate than concentration boundary conditions for our model electrolytic cell because chemical reactions at the electrodes determine the ionic fluxes at the boundaries, rather than fixed concentration reservoirs. Using flux boundary conditions introduces additional mathematical complexity in order to fully close the problem, distinguishing this method from that taken by \citet{wangSingularPerturbationSolutions2014}. We note that choosing constant flux boundary conditions also makes the analytic analysis of the steady state of this system for asymmetric valences more mathematically tractable compared to existing models with nonlinear interfacial reaction boundary conditions \autocite{richardsonTimedependentModellingAsymptotic2007,bazantCurrentVoltageRelationsElectrochemical2005}. Therefore, our minimal model provides a framework to examine, through an analytic approach, how asymmetric ion charges and fluxes alone impact the ionic concentrations and electric potential in a simple electrolytic cell. We provide a detailed analysis of the qualitative impacts of valence asymmetry on the system's steady state. \\

In an initial numerical study of this system, we observe a smooth transition between the near-equilibrium and strongly non-equilibrium regimes, characterised by a transition point at an intermediate current where the classical Debye-scale boundary layer vanishes and the electric field is approximately constant throughout the cell. This transition point depends on the ionic valence ratio of the electrolyte. To investigate this phenomenon further, we solve the PNP equations for a general electrolyte using asymptotic analysis \autocite{benderAdvancedMathematicalMethods2009}. We produce explicit analytic formulas for the composite asymptotic solutions for the ionic concentrations and electric potential in the \mbox{$z\R 2z$}, \mbox{$2z\R z$} and \mbox{$z\R z$} electrolyte. These explicit solutions can also be re-expressed for Dirichlet concentration boundary conditions, which then agree with the implicit solutions derived in  \citet{wangSingularPerturbationSolutions2014}.
Further, we give implicit solutions for the diffuse layer in the general asymmetric case and evaluate these solutions numerically (with explicit solutions for the bulk region). Finally, we fully characterise the observed transition behaviour in terms of ionic valences and fluxes and represent this on a phase diagram that can be used to predict the qualitative behaviour of the system.

\section*{Methods}
\subsection*{Problem Formulation}

\begin{figure}[t]
    \centering
    \begin{overpic}[width=0.75\linewidth,tics=5]{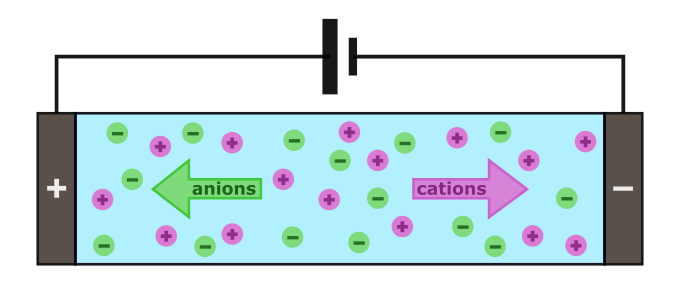}
    \put(10.5,1){0}
    \put(87,1){$\hat{L}$}
    \end{overpic}
    \caption{Schematic of a simple electrolytic cell showing the flux directions of the cations and anions in the dilute binary electrolyte and the relative positions of the electrodes. The anode at $\hat{x}=0$ has a higher potential than the cathode at $\hat{x}=\hat{L}$. For full generality, we allow both ionic species to undergo redox reactions at the electrodes. Cations are produced at the anode and consumed at the cathode, and vice versa for the anions.}
    \label{fig:placeholder}
\end{figure}

We create a minimal model to examine how asymmetric ion charges and fluxes fundamentally impact the ionic concentrations and electric potential in a simple electrochemical device.
We model an electrolytic cell in one dimension with parallel-plate electrodes at the boundaries and a dilute binary electrolyte $\text{M}_{\nu_{p}}\text{X}_{\nu_{n}}$, where $\nu_p$ and $\nu_n$ are the stoichiometric coefficients of the cation $\text{M}^{z_p+}$ and anion $\text{X}^{z_n-}$ respectively \autocite{hibbertIntroductionElectrochemistry1993}. Before the cell is turned on, the electrolyte has uniform concentration $\hat{C}$.  We consider the steady state of the system with no source or sink terms and no convection. We define the anode as the left electrode at $\hat{x}=0$ and the cathode as the right electrode at $\hat{x}=\hat{L}$, where $\hat{L}$ is the length of the electrolytic cell up to and including only the diffuse layer of the EDL at each electrode surface. See \cref{fig:placeholder} for a schematic. \\

We consider the non-equilibrium scenario where current flows through the electrolytic cell. The total electric current density in the cell $\hat{I}$, caused by the movement of ions, is 
\begin{equation}\label{dimcurrent}
    \hat{I}=Fz_p\hat{N}_p-Fz_n\hat{N}_n,
\end{equation}
 where $\hat{N_i} $ is the flux density of species $i=\{p,n\}$ where $p$ denotes cations and $n$ denotes anions, $z_i$ is the charge magnitude of species $i$, and $F$ is Faraday's constant \autocite{kontturiIonicTransportProcesses2015}. The flux at the boundaries arises due to the consumption and production of ions at the electrodes through redox reactions, i.e.~it is a Faradaic current\autocite{newmanElectrochemicalSystems2021}.  We simplify our notation by defining the ionic partial current density $\hat{I}_i$ carried by species $i$ via
\begin{align}
\refstepcounter{equation}%
\global\edef\CurrentFluxRelationNumber{\arabic{equation}}% store numeric parent
\phantomsection\label{CurrentFluxRelationAnchor}% create a hyperlink target at this spot
    \hat{I}_p:=Fz_p\hat{N}_p,\quad \hat{I}_n:=-Fz_{n}\hat{N}_n.
\tag{\CurrentFluxRelationNumber a,b}
\end{align}

This notational choice allows us to more easily frame our later discussion in terms of the total current density in the cell. It is common to see these quantities expressed in terms of transference or transport numbers. For reference, the transference (transport) numbers of the cations, $t_p$ and anions, $t_n$,  are defined as the fractions of the total current density carried by each species \autocite{kontturiIonicTransportProcesses2015},
\begin{align}
t_p := \frac{\hat I_p}{\hat I}, \quad
t_n := \frac{\hat I_n}{\hat I}.
\refstepcounter{equation}
\tag{\theequation a,b}
\end{align}

The steady-state dimensional PNP system we need to solve for the cation concentration $\hat{p}$, anion concentration $\hat{n}$ and electric potential $\hat{\phi}$ is given by
\begin{subequations}\label{dimensionalPNP}
\begin{align}
    \frac{\mathrm{d}}{\mathrm{d}\hat{x}} \left(-\hat{D}_p \frac{\mathrm{d}\hat{p}}{\mathrm{d}\hat{x}} -\frac{z_p \hat{D}_p F}{R\hat{T}}\hat{p} \frac{\mathrm{d}\hat{\phi}}{\mathrm{d}\hat{x}}  \right)&=0, \label{Nernst-Planck Continuity p} \\
    \frac{\mathrm{d}}{\mathrm{d}\hat{x}} \left(-\hat{D}_n \frac{\mathrm{d}\hat{n}}{\mathrm{d}\hat{x}} +\frac{z_n \hat{D}_nF}{R\hat{T}}\hat{n} \frac{\mathrm{d}\hat{\phi}}{\mathrm{d}\hat{x}} \right)&=0, \label{Nernst-Planck Continuity n} \\
    \hat{\epsilon} \frac{\mathrm{d}^2\hat{\phi}}{\mathrm{d}\hat{x}^2}  =-F \left(z_p \hat{p}-z_n \hat{n}\right)&, \label{Poisson1D}
\end{align}
\end{subequations}

where $F$ is Faraday's constant, $R$ is the ideal gas constant, $\hat{T}$ is the temperature, $\hat{\epsilon}$ is the dielectric permittivity, and $\hat{D}_i$ is the diffusion coefficient of species $i$ \autocite{newmanElectrochemicalSystems2021}. The PNP system is composed of the steady-state continuity equations for ionic concentration, \eqref{Nernst-Planck Continuity p} and \eqref{Nernst-Planck Continuity n}, where the fluxes are given by the Nernst--Planck flux equations, and the Poisson equation \eqref{Poisson1D}, which describes how the potential responds to regions of space-charge.\\

Our minimal model has the following associated conditions:
\begin{subequations}
\label{eq: BC dim}
    \begin{alignat}{2}
        \left. \left(\hat{D}_p\frac{\mathrm{d} \hat{\pOG}}{\mathrm{d} \hat{x}}+\frac{\hat{D}_pF }{R \hat{T}} z_p\hat{\pOG} \frac{\mathrm{d}\hat{\phi}}{\mathrm{d} \hat{x}}\right)\right|_{\hat{x}=0,\hat{L}}&=-\frac{ \hat{I}_p}{Fz_p}  , \label{FluxBC1} \\
\left.\left(\hat{D}_n\frac{\mathrm{d} \hat{\nOG}}{\mathrm{d} \hat{x}}-\frac{\hat{D}_nF}{R \hat{T}}  z_n\hat{\nOG} \frac{\mathrm{d}\hat{\phi}}{\mathrm{d} \hat{x}}\right)\right|_{\hat{x}=0,\hat{L}}&=\frac{ \hat{I}_n}{Fz_n},\label{FluxBC2}\\
\hat{\phi}(0)=\hat{V}>0, \qquad \hat{\phi}(\hat{L})&=0, 
\refstepcounter{equation}
\tag{\theequation,d} \label{BCsPotential}
\\
\dint{0}{\hat{L}}{\pOG}{x}=\nu_p\hat{C}, \quad \dint{0}{\hat{L}}{\nOG}{x}&=\nu_n\hat{C} .
\refstepcounter{equation}
\refstepcounter{equation}
\tag{\theequation,f}
\label{BC:phi dimensional}
    \end{alignat}
\end{subequations}
The flux boundary condition \eqref{FluxBC1} represents redox reactions occurring at the electrodes, producing cations at the anode and consuming cations at the cathode with flux $\hat{I}_p/{Fz_p}$, and likewise \eqref{FluxBC2} describes anions being produced at the cathode and consumed at the anode. The potential boundary conditions \eqref{BCsPotential} define the potential drop $\hat{V}$ from the far edges of one diffuse layer to the other (a few ions' distance away from the electrodes into the cell interior), with the anode having a higher potential than the cathode. Finally, \eqref{BC:phi dimensional} describes mass conservation throughout the system to follow global electroneutrality based on the electrolyte's initial concentration. \\

We note that, for notational convenience, we take $\hat{I}_p,\hat{I}_n\ge0$ and introduce a negative sign in \eqref{FluxBC1} to ensure both ion species have flux in the expected directions according to our choice of electrode configuration \autocite{newmanElectrochemicalSystems2021}. Despite acting on both boundaries, mass conservation means that \eqref{FluxBC1} only represents one independent boundary condition at steady state (with similar for \eqref{FluxBC2}). Hence, \eqref{eq: BC dim} represent six boundary conditions for the three second-order ODEs in \eqref{dimensionalPNP}, as required. \\
 
We note that our potential and flux boundary conditions are `simple' modelling choices. We have already discussed that constant flux boundary conditions (and  mass-conservation) simplify the mathematics relative to nonlinear interfacial boundary conditions like Butler-Volmer reaction kinetics\autocite{newmanElectrochemicalSystems2021}. This makes the asymmetric valence calculations more mathematically tractable, enabling analytic expressions for certain valence ratio values. Choosing Dirichlet boundary conditions for the potential \eqref{BCsPotential} is effectively employing the Gouy--Chapman limit to ignore the potential drop in the Stern layer of the EDL \autocite{vansoestbergenDiffusechargeEffectsTransient2010,bazantCurrentVoltageRelationsElectrochemical2005}. Common approaches to modelling the Stern layer invoke more complex boundary conditions and assumptions as the continuum approximation breaks down \autocite{bazantCurrentVoltageRelationsElectrochemical2005,vansoestbergenDiffusechargeEffectsTransient2010}. Since detailed dynamics at the electrodes are not the focus of this work, we choose to keep this model as simple as possible.  We also note that, for our mathematical analysis, the potential can be translated without loss of generality because the PNP system depends only on potential gradients. 

\subsection*{Non-dimensionalisation} \label{NondimensionalisationOverall}
We introduce the following scalings to non-dimensionalise the problem: 
\begin{align}
\refstepcounter{equation}
\global\edef\NondimNumber{\arabic{equation}}% store numeric parent
\phantomsection\label{Nondim}% create a hyperlink target at this spot
    &\hat{\pOG}=\hat{C}\nu_{p} \pOG , \quad \hat{\nOG}=\hat{C}\nu_{n}\nOG,\quad \hat{x}=\hat{L}x, \quad \hat{\potOG}=\frac{R\hat{T}}{F z_n}\potOG,\quad \hat{V}=\frac{R\hat{T}}{F z_n}V, \nonumber \\
    &\hspace{14mm}  \hat{I}_p=\frac{F\hat{C}\hat{D}_{p}z_{p}\nu_{p}}{\hat{L}} I_p,\quad \hat{I}_n=\frac{F\hat{C}\hat{D}_{n}z_{n}\nu_{n}}{\hat{L}} I_n.
    \tag{\NondimNumber a--g}
\end{align}
Here, the potentials are scaled by the thermal voltage divided by the anion valence. The ion concentrations are scaled by their stoichiometric coefficients and the electrolyte's initial concentration, so that electroneutrality at a point implies that the dimensionless cation and anion concentrations are equal. We now define a reference concentration using the electroneutrality properties of the binary electrolyte,
\begin{equation}
    \mathcal{C}=z_p\nu_p \hat{C}=z_n\nu_n \hat{C}.
\end{equation}
Hence, we can define the dimensionless small parameter,
\begin{align}
    \eps=\frac{1}{\hat{L}} \sqrt{  \frac{\hat{\epsilon} R\hat{T}}{F^2z_n\mathcal{C}} }\ll1, \label{EpsDef}
\end{align}
which is a ratio of the characteristic length of the diffuse layer in this model to the length of the cell. We note that $\eps$ is related to the classic Debye length\autocite{newmanElectrochemicalSystems2021} $\lambda$ via
\begin{equation}
    \hat{L} \eps =\lambda \sqrt{1+\rr},
    \end{equation}
    where we define the ion valence ratio, \begin{equation}
    r=\frac{z_p}{z_n} \in \mathbb{Q}_{>0}.\label{ratio}
\end{equation}
We emphasize that when considering asymmetric valence binary electrolytes, it is more mathematically convenient to use the characteristic length $\hat{L} \eps$ for the diffuse layer than $\lambda$.\\

Applying the scalings \hyperref[Nondim]{[\NondimNumber]}, we can now express the dimensionless Poisson--Nernst--Planck system of equations in terms of the ion valence ratio $\rr$ and the small parameter $\eps$,
\begin{subequations}\label{FluxGovEqns}
\begin{align}
\dd{}{x}\left(\frac{\mathrm{d} \pOG}{\mathrm{d} x}+ r\pOG\frac{\mathrm{d} \potOG}{\mathrm{d} x}\right)&=0\label{numPEqn},\\
\dd{}{x}\left(\frac{\mathrm{d} \nOG}{\mathrm{d} x}- \nOG\frac{\mathrm{d} \potOG}{\mathrm{d} x}\right)&=0\label{numNEqn}, \\
\eps^2\frac{\mathrm{d}^2\potOG}{\mathrm{d} x^2} =-\pOG+\nOG&, \label{numPhiEqn}
\end{align}
\end{subequations}

with the following dimensionless conditions,
\begin{subequations}\label{FluxGovEqnsBCs}
\begin{alignat}{2}
\left.\left(\frac{\mathrm{d} \pOG}{\mathrm{d} x}+ r\pOG\frac{\mathrm{d} \potOG}{\mathrm{d} x}\right)\right|_{x=0,1}&=-\Ip \label{NumBC1},\\
\left.\left(\frac{\mathrm{d} \nOG}{\mathrm{d} x}- \nOG\frac{\mathrm{d} \potOG}{\mathrm{d} x}\right)\right|_{x=0,1}&=\In \label{NumBC2}, \\
\potOG(0)= V, \quad  \quad \potOG(1)&=0. \refstepcounter{equation}
\tag{\theequation,d}\\
\dint{0}{1}{p}{x}=1, \quad \dint{0}{1}{n}{x}&=1 .
\refstepcounter{equation}
\refstepcounter{equation}
\tag{\theequation,f}
\end{alignat}
\end{subequations}
Since we have not included any explicit interfacial reaction terms in our flux boundary conditions, instead opting for constant flux terms $I_p$ and $I_n$, the system must satisfy \eqref{NumBC1} and \eqref{NumBC2} through transport mechanisms alone (diffusion and electromigration). Therefore, the scope of our minimal model is to analyse how valence asymmetry impacts ionic transport behaviour in isolation. \\

Integrating [\ref{FluxGovEqns}a,b] once and applying the boundary conditions [\ref{FluxGovEqnsBCs}a,b], this system simplifies to 
\begin{subequations}\label{FluxGovEqnsInt}
\begin{align}
\frac{\mathrm{d} \pOG}{\mathrm{d} x}+ \rr\pOG\frac{\mathrm{d} \potOG}{\mathrm{d} x}&=-I_p\label{NEqn},\\
\frac{\mathrm{d} \nOG}{\mathrm{d} x}- \nOG\frac{\mathrm{d} \potOG}{\mathrm{d} x}&=I_n\label{PEqn}, \\
\eps^2\frac{\mathrm{d}^2\potOG}{\mathrm{d} x^2} &=-\pOG+\nOG, \label{Poisson}
\end{align}
\end{subequations}
with remaining conditions,
\begin{align}
\refstepcounter{equation}
\global\edef\BCsNumber{\arabic{equation}}% store numeric parent
\phantomsection\label{FluxBCsInt}% create a hyperlink target at this spot
\potOG(0)=V,&\quad \potOG(1)=0, \quad \dint{0}{1}{n}{x}=1,\quad \dint{0}{1}{p}{x}=1.
\tag{\BCsNumber a--d}
\end{align}
Note that the mass-conservation conditions \hyperref[FluxBCsInt]{[\BCsNumber c,d]} arise as a consequence of the duplication of the same flux condition at each boundary in [\ref{FluxGovEqnsBCs}a,b]. The total mass for the cations and anions follows from the assumed initial concentration of the electrolyte and the concentration scalings [\NondimNumber a,b].

\subsection*{Numerical Simulations}\label{Sec:Numerics}
We start by investigating numerically the impact of changing the valence ratio $\rr$ on the behaviour of this model. 
We model the time-dependent version of \eqref{FluxGovEqns} (re-instating the appropriate concentration time derivatives on the right-hand side of the continuity equations \eqref{numPEqn} and \eqref{numNEqn} \autocite{newmanElectrochemicalSystems2021}) using a method of lines approach and evolve the system until it reaches steady state. We solve \eqref{numPhiEqn} as an elliptic constraint at each time-step. The time-dependent equations are solved using Python \autocite{Python} and the packages \texttt{scipy} \autocite{SciPy} and \texttt{numpy} \autocite{NumPy}. We use \texttt{scipy.integrate.solve\_ivp} with the BDF method and the constraint is enforced by solving with \texttt{scipy.sparse.linalg.spsolve}.  Code is available in a GitHub repository\autocite{georgina_ryan_2026_19695204}.\\

\begin{figure}
    \centering
        \begin{overpic}[width=\textwidth,tics=5]{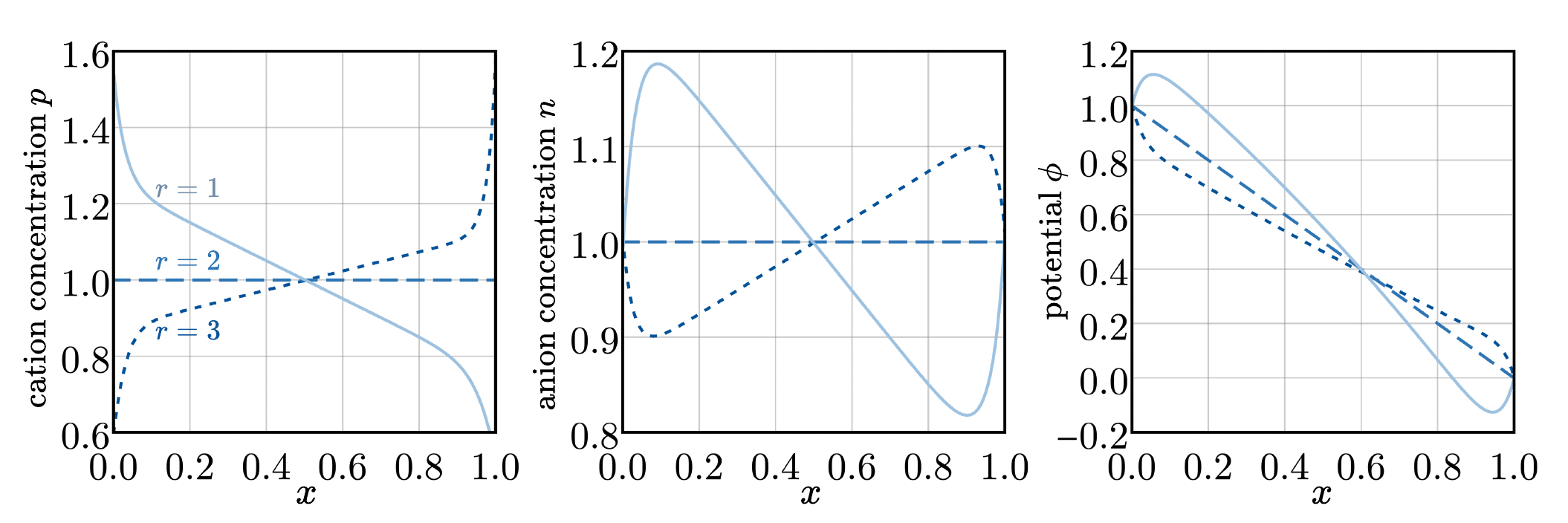}
    \end{overpic}
    \caption{Numerical solutions for the ion concentrations $\pOG$, $\nOG$ and the electric potential $\potOG$ respectively of \eqref{FluxGovEqnsInt} and \hyperref[FluxBCsInt]{[\BCsNumber]} for fixed $I_p=2$, $I_n=1$, $V=1$, and $\eps=0.05$, shown for $\rr=1$ (light solid), $\rr=2$ (medium dashed), and $\rr=3$ (dark dotted). \textbf{a.} Cation concentration $p$.
\textbf{b.} Anion concentration $n$.
\textbf{c.} Electric potential $\phi$.}
    \label{fig:numerical initial results}
\end{figure}

In \cref{fig:numerical initial results}, we consider the spatial dependence of cation concentration $p$, anion concentration $n$, and potential $\phi$ as we vary the valence ratio $\rr=\{1,2,3\}$ with fixed fluxes, and see that this can induce very different qualitative behaviour. Note that \cref{fig:numerical initial results} has $I_p=2$, $I_n=1$; the distinct qualitative behaviours demonstrated in these plots are achievable by an electrolyte of any valence ratio under the appropriate fluxes. \\

Firstly, for the $\rr=3$ electrolyte,  the boundary layers in the  concentration and potential are qualitatively similar to the predictions of Gouy--Chapman theory\autocite{bardElectrochemicalMethodsFundamentals2022}. Namely, the concentration of the cations accumulates at the cathode ($x=1$) and the concentration of anions accumulates at the anode ($x=0$). Additionally, the potential decreases monotonically across the domain. As Gouy--Chapman is strictly an equilibrium theory and these numerical results are for a non-zero current, we refer to these qualitative behaviours as a `near-Gouy--Chapman' regime. \\

Secondly, for the $\rr=2$ electrolyte, there is a uniform unit spatial concentration, and the potential curve is a linear function between $\phi(0)=1$ and $\phi(1)=0$. That is, there are no boundary layer contributions to the solutions. We note \citet{wangSingularPerturbationSolutions2014} observed no boundary layers if they chose an electroneutral setup for a tertiary electrolyte with Dirichlet concentration boundary conditions. Therefore, there is precedent for observing no boundary layers in the results of the PNP equations. \\

Finally, for the $\rr=1$ electrolyte, the behaviour of the concentrations near the boundaries appears to be the opposite of that observed for the $\rr = 3$ electrolyte. Specifically, the cation concentration is depleted at the cathode and the anion concentration is depleted at the anode. This is the characteristic condition of approaching the limiting current, as ions are consumed at the electrode faster than they can be replenished from bulk ion transport \autocite{bardElectrochemicalMethodsFundamentals2022}.  In this case, the potential becomes non-monotonic, indicating a change in the direction of the electric field throughout the cell.  \\

Two limiting cases of our minimal model are Gouy--Chapman theory\autocite{gouyConstitutionChargeElectrique1910,chapmanLIContributionTheory1913} at zero current and limiting current behaviour\autocite{newmanElectrochemicalSystems2021} for high currents, and therefore we expect a transition between these two states at intermediate currents.
Our numerical observations indicate that the transition between these states as the current increases must depend on the valence ratio. We now use asymptotic analysis to understand, characterise, and quantify this transition behaviour, focusing on the role of the valence ratio.

\subsection*{Asymptotic analysis}
We conduct a boundary layer analysis\autocite{benderAdvancedMathematicalMethods2009} of our problem \eqref{FluxGovEqnsInt} with conditions \hyperref[FluxBCsInt]{[\BCsNumber]} to solve for leading order in $\eps$ approximations for the ion concentrations $p,n$ and potential $\phi$. We will find that the domain comprises three regions: the electroneutral bulk region (the `outer region' in the language of matched asymptotic expansions) and the two diffuse layer regions (boundary layers) near $x=0$ and $x=1$. After solving in each domain, we use asymptotic matching between these three regions to define the global approximations to $p,n,\phi$, which are accurate in the physically relevant limit $\eps \to 0$.  

\subsubsection*{Electroneutral Bulk}
We introduce asymptotic expansions in powers of the small parameter $\eps\ll 1$ in the outer region, away from the boundaries at $x=0$ and $x=1$: 
\begin{subequations}\label{concDefsSymm}
\begin{align}
\potOG(x) &= \pot(x)+\eps \potOG_1(x)+O(\eps^2), \label{PotSeriesSymm}\\
\pOG(x) &= \p(x)+\eps p_1(x)+O(\eps^2), \label{pSeriesSymm}\\
\nOG(x) &= \n(x)+\eps n_1(x)+O(\eps^2). \label{nSeriesSymm}
\end{align}
\end{subequations}

We introduce two parameters to simplify our calculations: the weighted total current,
\begin{equation}\label{total current def}
\II=\frac{\Ip+\In}{1+\rr}\ge0,
\end{equation}
%an expression related to the sum of the valences,
%\begin{equation}
    %\K=1+\rr=\frac{z_p+z_n}{z_n},
%\end{equation}
and the weighted current imbalance,
\begin{equation}
\B=\frac{I_p-\rr I_n}{2 (1+\rr)}.\label{B definition}
\end{equation}
Since both $\II$ and $\B$ involve the valence ratio $\rr$, any system behaviour that depends on these parameters implicitly depends on the valence ratio.\\

We first solve for the leading-order potential $\pot$ and concentrations $\p,\n$ in the outer region. Substituting \eqref{concDefsSymm} into the Poisson equation \eqref{Poisson}, we obtain the leading-order solutions,
\begin{equation} \label{c0 equal}
\p(x)=\n(x):=\C(x).
\end{equation}
To determine $\C$, we first substitute \eqref{c0 equal} into the Nernst--Planck flux equations \hyperref[FluxGovEqnsInt]{[\ref{FluxGovEqnsInt}ab]}. Then, adding \eqref{PEqn} to $\rr$ times \eqref{NEqn}, we find,
\begin{equation}
\frac{\mathrm{d }\C}{\mathrm{d}x}=-2\B. \label{DEc0}
\end{equation}
Integrating \eqref{DEc0} and applying the leading-order mass-conservation conditions \hyperref[FluxBCsInt]{[\BCsNumber c,d]},
we obtain
\begin{equation}
\C(x)=1-2\B \left(x-\frac{1}{2}\right). \label{Cout2Flux}
\end{equation}
%We could hold off applying the mass-conservation condition at this stage, but the end result is the same and this is simpler notationally. \\
Next, by subtracting \eqref{NEqn} from \eqref{PEqn}, we obtain a differential equation for the leading-order potential $\pot$,
\begin{equation}
\frac{\mathrm{d }\pot}{\mathrm{d}x}=-\frac{\II}{\C}.\label{PotentialDE}
\end{equation}
Therefore, substituting \eqref{Cout2Flux} into \eqref{PotentialDE} and integrating, we find
\begin{equation}
\pot(x)=\frac{\II}{2\B } \log\left( -2\B x+1+\B \right)+\Kc, \label{PhiOut1Asym}
\end{equation}
where $\Kc$ is a constant that we will determine below by matching to the boundary layers.

\subsubsection*{Left Diffuse Layer}
We now scale into the boundary layer variable $x=\eps \xL$ where $\xL=O(1)$. We will only require the leading-order solution in the boundary layers so, for notational brevity, we consider leading-order contributions in $\eps$ in the left boundary layer: cation concentration $p(x)\sim\pL(\xL)$, anion concentration $n(x)\sim\nL(\xL)$, and potential $\phi(x)\sim\potL(\xL)$. Substituting these leading-order terms into the PNP equations \eqref{FluxGovEqns} with the boundary layer variable $\xL$, we obtain the governing equations:
\begin{subequations}\label{LeftFluxEqns}
\begin{align}
\dd{\pL}{\xL}+\rr\pL \dd{\potL}{\xL}&=0, \label{Leftpeqn2b}\\
\dd{\nL}{\xL}-\nL \dd{\potL}{\xL}&=0, \label{Leftneqn2b} \\
\dd{^2\potL}{\xL^2}&=-\pL+\nL. \label{LeftPotEqn2}
\end{align}
\end{subequations}
For notational convenience, we write $p(0)=p_L$, $p(1)=p_R$, $n(0)=n_L$ and $n(1)=n_R$, where $p_L$, $p_R$, $n_L$, $n_R>0$ are the four remaining degrees of freedom in the problem. 
Since the asymptotic magnitude of the solutions does not vary between the outer and boundary layer solutions, we may determine the matching conditions between the left boundary layer and the outer region by equating the outer limit of the boundary layer solution with the inner limit of the outer solution,
\begin{align}
\refstepcounter{equation}%
\global\edef\GenMatchingNumber{\arabic{equation}}% store numeric parent
\phantomsection\label{GenMatchingAnchor}% create a hyperlink target at this spot
\lim_{ \xL \to \infty } \potL\left(\xL\right)=\lim_{ x \to 0 } \pot(x), \quad \lim_{ \xL \to \infty } \pL\left(\xL\right)=\lim_{ x \to 0 } \C(x), \quad \lim_{ \xL \to \infty } \nL\left(\xL\right)=\lim_{ x \to 0 } \C(x).
\tag{\GenMatchingNumber a--c}
\end{align}

Therefore, the left boundary layer has the following matching conditions,
\begin{subequations}\label{MatchingBCsLeft}
\begin{align}
\potL\left(\infty\right)&=\phi_0(0)=\frac{\II}{2\B}\log\left( 1+\B \right)+\Kc ,\label{Phi0ConditionL2}\\
\pL(\infty)&=\nL(\infty)=1+\B ,\label{p0ConditionL2}
\end{align}
\end{subequations}
and boundary conditions,
\begin{align}
\refstepcounter{equation}%
\global\edef\LeftBCsNumber{\arabic{equation}}% store numeric parent
\phantomsection\label{LeftBCsAnchor}% create a hyperlink target at this spot
\potL(0)=V, \quad &\pL(0)=p_L, \quad \nL(0)=n_L,
\tag{\LeftBCsNumber a--c}
\end{align}
where $p_L$, $n_L$, $\Kc$ are constants to be determined. \\

Solving \eqref{LeftFluxEqns} with boundary conditions \hyperref[LeftBCsAnchor]{[\LeftBCsNumber]}, we derive Boltzmann relations valid at leading order in the left boundary layer region,
\begin{align}
\refstepcounter{equation}%
\global\edef\BoltzmannLeftNumber{\arabic{equation}}% store numeric parent
\phantomsection\label{BoltzmannLeftAnchor}% 
\pL(\xL)&=p_L\exp(\rr(V-\potL(\xL))), \quad \nL(\xL)=n_L\exp(\potL(\xL)-V).
\tag{\BoltzmannLeftNumber a,b}
\end{align}

It is now convenient to use the matching conditions \eqref{MatchingBCsLeft} 
to write the integration constant $\Kc$ and $p_L$ in terms of $n_L$. This step reduces our problem to three remaining degrees of freedom and simplifies \eqref{LeftPotEqn2}. Multiplying together $\pL$ and $\nL$ and dividing $\pL$ by $\nL$ from \hyperref[BoltzmannLeftAnchor]{[\BoltzmannLeftNumber]} after applying the matching conditions \eqref{MatchingBCsLeft} yields
\begin{subequations}
    \begin{align}
        p_L &=\frac{(1+\B)^2}{n_L}\exp[(\rr-1)(\pot(0)-V)],\label{pLDef2B}\\
        \pot(0) &=V+\frac{1}{1+\rr}\log\left( \frac{p_{L}}{n_{L}} \right). \label{potL0}
    \end{align}
\end{subequations}

Therefore, substituting \eqref{potL0} into \eqref{pLDef2B}, we find
\begin{align}
\refstepcounter{equation}%
\global\edef\pLAsymDefNumber{\arabic{equation}}% store numeric parent
\phantomsection\label{pLAsymDefAnchor}% 
p_{L}=\frac{\left( 1+\B \right)^{1+\rr}}{n_{L}^{\rr}}, \quad \pot(0) =V+\log\left( \frac{1+\B}{n_{L}} \right).
\tag{\pLAsymDefNumber a,b}
\end{align}
The physical restriction that $p_L>0$ enforces $\B>-1$. We can now express $\Kc$ in terms of $n_L$ by equating $\pot(0)$ from \eqref{potL0}, which was derived using information from the boundary layer, with the definition of $\pot(0)$ from the outer region \eqref{Phi0ConditionL2}. We therefore find that the outer solution $\pot(x)$ \eqref{PhiOut1Asym} can be written as
\begin{equation}
   \pot(x)=V+\frac{\II}{2\B } \log\left( -2\B x+1+\B \right)-\frac{\II}{2\B}\log\left( 1+\B \right) + \log \left(\frac{1+\B}{n_L}\right). \label{fullOuterPotential2}
\end{equation}
This outer solution \eqref{fullOuterPotential2} still contains one degree of freedom, $n_L$, so we still do not have enough information to close the outer problem. There are now three degrees of freedom remaining in the system: $n_L$, $n_R$ and $p_R$.\\

We now use the relationship between $p_L$ and $n_L$ \hyperref[pLAsymDefAnchor]{[\pLAsymDefNumber a]} to significantly simplify the differential equation for the left boundary layer \eqref{LeftPotEqn2}. First, we substitute $P,N$ \hyperref[BoltzmannLeftAnchor]{[\BoltzmannLeftNumber]} and $p_L$ \hyperref[pLAsymDefAnchor]{[\pLAsymDefNumber a]} into the Poisson equation \eqref{LeftPotEqn2} to obtain 
\begin{equation}
\dd{^2\potL}{\xL^2}=-\frac{\left( 1+\B \right)^{1+\rr}}{n_{L}^{\rr}}\exp(r(V-\potL(\xL)))+n_L\exp(\potL(\xL)-V). \label{SubstitutedLeftEqn}
\end{equation}
We then multiply \eqref{SubstitutedLeftEqn} by $\mathrm{d}\potL/\mathrm{d}\xL$ and integrate. The integration constant is fixed by the matching conditions \eqref{MatchingBCsLeft}, the definition of $\phi_{0}(0)$ \hyperref[pLAsymDefAnchor]{[\pLAsymDefNumber b]}, and the implication of the matching conditions that $\potL'(\infty )=0$.
Hence, we find that 
\begin{align}
\frac{1}{2}\left(\frac{\mathrm{d}\potL}{\mathrm{d}\xL}\right)^2= \frac{\left( 1+\B \right)^{1+\rr}}{\rr n_{L}^{\rr}}\exp(\rr(V-\potL(\xL)))+n_{L}\exp(\potL(\xL)-V)-\frac{1+\rr}{\rr} \left( 1+\B \right).  
\label{diffeqn}
\end{align}

To proceed, it is helpful to convert the nonlinear functions of $\potL$ in \eqref{diffeqn} into powers of a transformed variable. Therefore, we make the substitution \begin{equation} \label{mSub}
     \mL(\xL)=\exp((\potL(\xL)-V)/2).
\end{equation}  
Under this substitution, the relevant boundary and matching conditions from \eqref{MatchingBCsLeft} become
\begin{align} 
\refstepcounter{equation}%
\global\edef\mBCsNumber{\arabic{equation}}% store numeric parent
\phantomsection\label{mBCsAnchor}% create a hyperlink target at this spot
    \mL(0)=1, \quad \mL(\infty)=\exp((\potOG(0)-V)/2)=\tl,
\tag{\mBCsNumber a,b}
\end{align} where 
\begin{equation}
    \tl=\sqrt{\frac{1+\B}{n_L}}>0.\label{alpha Asym def}
\end{equation} 
 With the substitution \eqref{mSub} and taking the square root, \eqref{diffeqn} becomes
\begin{align}
\frac{\mathrm{d}\mL}{\mathrm{d}\xL}= \text{sign}(\alpha_L-1)\sqrt{\frac{n_L}{2}}\sqrt{\frac{\tl^{2(1+\rr)}}{\rr}\mL^{2-2 \rr}+ \mL^{4}- \frac{1+\rr}{\rr} \tl^2 \mL^2}. \label{Gen LBL eqn}
\end{align}
\\
Therefore, the implicit analytic solution in this boundary layer is
\begin{equation}\label{ImplicitLeft}
\xL
= \text{sign}(\alpha_L-1)\,
\sqrt{\frac{2}{n_L}}
\int_{1}^{\mL(\xL)}
\left(
\frac{\tl^{2(1+\rr)}}{\rr} s^{\,2-2 \rr}
+ s^{4 }
-\frac{1+\rr}{\rr} \tl^2 s^{\,2}
\right)^{-1/2}
\,\mathrm{d}s,
\end{equation}
using \hyperref[mBCsAnchor]{[\mBCsNumber a]}.  At this stage, there are still three remaining degrees of freedom in the problem: $n_L, p_R, n_R$. \\

In general, \eqref{ImplicitLeft} is an elliptic or hyperelliptic integral\autocite{byrdHandbookEllipticIntegrals1971} and so cannot be expressed in terms of elementary functions. However, in the cases where $\rr=\{1/2, 1, 2\}$, the symmetries of the integrand mean that we can solve \eqref{ImplicitLeft} in terms of elementary functions. Recall that these are also the cases with known elementary solutions for the single-electrode Poisson--Boltzmann problem \autocite{gouyConstitutionChargeElectrique1910,andriettiExactSolutionUnidimensional1976}. We have been unable to express \eqref{ImplicitLeft} in terms of elementary functions for other physically realistic values for $\rr$ (i.e. corresponding to $z_p,z_n \in \{1,2,3,4\}$). Therefore, for general asymmetric electrolytes, we proceed with the implicit solution \eqref{ImplicitLeft} or solve \eqref{Gen LBL eqn} numerically in Mathematica\autocite{Mathematica} using \texttt{NDSolve} with the \texttt{Stiffness Switching} method.\\

To motivate why it seems as though only the $\rr=\{1/2, 1, 2\}$ cases can be expressed in terms of elementary functions, we note that \eqref{ImplicitLeft} is of the form $\int \mathrm{d}s/\sqrt{P(s)}$ where $P(s)$ is a finite Puiseux series\autocite{raabIntegrationFiniteTerms2022} (i.e. a power series that may include negative or fractional exponents). Liouville's theorem on elementary integration dictates that integrals where $P(s)$ is a cubic or higher-order polynomial are not representable by elementary functions \autocite{raabIntegrationFiniteTerms2022}. The exception is if there is a repeated root in $P(s)$ such that, after factorisation, the irreducible polynomial remaining under the square root is quadratic or linear. In this case, the integral is elementary \autocite{byrdHandbookEllipticIntegrals1971}.  This exception covers $\rr=1/2$ and $\rr=1$, for which $P(s)$ takes the form of a quartic polynomial with repeated roots.\\

The $\rr=2$ case has a negative exponent in $P(s)$, so we cannot directly apply the same argument. However, an alternative non-dimensionalisation (where we scale by $z_p$ instead of $z_n$ in \hyperref[Nondim]{[\NondimNumber d,e]} and \eqref{EpsDef}) would lead to a scenario where we can use this argument. Briefly, the alternative non-dimensionalisation leads to the scenario where, after writing the governing differential equation in the left boundary layer in terms of $p_L$, the resulting equation has a cubic form for $P(s)$ with a repeated root, and hence has an elementary solution. Since this alternative non-dimensionalisation represents the same physical problem, the solutions under both non-dimensionalisations will be related by a simple change of variables. Hence, we expect an elementary solution in the $\rr=2$ case with the scalings in \hyperref[Nondim]{[\NondimNumber]}. 

\subsubsection*{Right Diffuse Layer}
We briefly outline the solution in the right boundary layer. Here, we transform to a boundary layer variable $\xR=O(1)$ where $x=1-\eps \xR$. We introduce notation to represent the leading-order contributions for the cation concentration $p(x)\sim \pR(\xR)$, anion concentration $n(x)\sim\nR(\xR)$, and potential $\potOG(x) \sim \potR(\xR)$. Following an equivalent matching procedure to \eqref{LeftFluxEqns}--\hyperref[pLAsymDefAnchor]{[\pLAsymDefNumber a]}  yields 
\begin{align}
\refstepcounter{equation}%
\global\edef\RightPotentialNumber{\arabic{equation}}% store numeric parent
\phantomsection\label{RightPotentialAnchor}% create a hyperlink target at this spot
    p_R=\frac{(1-\B)^{(1+\rr)}}{n_R^{\rr}}, \quad \quad\potR(\infty)=\pot(1)=\frac{1}{1+\rr}\log\left( \frac{p_{R}}{n_{R}} \right)=\log\left(\frac{1-\B}{n_R}\right).\tag{\RightPotentialNumber a,b}
\end{align}
The restrictions that $p_R>0$ in \hyperref[RightPotentialAnchor]{[\RightPotentialNumber a]} and $n_L>0$ from \hyperref[pLAsymDefAnchor]{[\pLAsymDefNumber a]}  necessitates that $|\B|<1$. \\

There are now two degrees of freedom remaining: $n_L$ and $n_R$. We can remove one degree of freedom by equating our representation of $\potOG(1)$ from the outer solution \eqref{fullOuterPotential2} with the representation derived from the right boundary layer \hyperref[RightPotentialAnchor]{[\RightPotentialNumber a]}. Hence, 
\begin{equation}
    n_L =\NN^2 n_R \quad \text{ where } \quad \NN=\exp(V/2)\left(\frac{1-\B}{1+\B}\right)^{\frac{\II }{4\B} -\frac{1}{2}}.
   \label{MultiplyingMatchRight2B}
\end{equation}
Importantly, our solution procedure shows that information from each boundary layer ($n_L, n_R$) propagates through the outer region to the other boundary layer, affecting the solutions therein. \\

We proceed by solving the Poisson equation \eqref{Poisson} in the right inner region,
\begin{equation}
\dd{^2\potR}{\xR^2}=-\frac{\left( 1-\B \right)^{(1+\rr)}}{n_{R}^{\rr}}\exp(-\rr\potR(\xR))+n_R\exp(\potR(\xR)), \label{SubstitutedRightEqn}
\end{equation}

using an analogous method to the left inner region (from \eqref{SubstitutedLeftEqn} to \eqref{Gen LBL eqn}), now expressing all constants in terms of $n_R$. This process yields the differential equation 
\begin{align}
\frac{\mathrm{d}\mR}{\mathrm{d}\xR}= \text{sign}(\alpha_R-1)\sqrt{\frac{n_R}{2}}\sqrt{\frac{\tr^{2(1+\rr)}}{\rr}\mR^{2-2 \rr}+\mR^{4 }-\frac{1+\rr}{\rr} \tr^2\mR^2} ,\label{Gen RBL Eqn}
\end{align}
where 
\begin{align}
\refstepcounter{equation}%
\global\edef\SubRightNumber{\arabic{equation}}% store numeric parent
\phantomsection\label{SubRightAnchor}% create a hyperlink target at this spot
    \mR(\xR)=\exp(\potR(\xR)/2), \quad \tr=\sqrt{\frac{1-\B}{n_R}} ,\tag{\SubRightNumber a,b}
\end{align}
with boundary and matching conditions,
\begin{equation} \label{M BCs}
    \mR(0)=1, \quad \mR(\infty)=\exp(\potOG(1)/2)=\tr.\refstepcounter{equation}\tag{\theequation \text{a,b}}
\end{equation}

The implicit solution in this boundary layer is therefore,
\begin{equation}\label{ImplicitRight}
\xR
= \text{sign}(\alpha_R-1)\,
{\sqrt\frac{2}{n_R}}
\int_{1}^{\mR(\xR)}
\left(
\frac{\tr^{2(1+\rr)}}{\rr} s^{\,2-2\rr}
+s^{4}
-\frac{(1+\rr)}{\rr} \tr^2 s^{\,2}
\right)^{-1/2}
\,\mathrm{d}s .
\end{equation}

At this stage, there is one remaining degree of freedom, $n_R$, so the problem is not closed. We now derive the necessary closure condition to fix $n_R$.  

\subsubsection*{Closing the problem}
\label{Sec:ClosingTheProblem}
By deriving implicit solutions in the boundary layers, \eqref{ImplicitLeft} and \eqref{ImplicitRight}, we have determined implicit solutions for $\potL(\xL)$ and $\potR(\xR)$ by substituting back into
\begin{align} 
\refstepcounter{equation}%
\global\edef\PotDefNumber{\arabic{equation}}% store numeric parent
\phantomsection\label{potential defs}% create a hyperlink target at this spot
    \potL(\xL)=V+2\log(\mL(\xL)), \quad \potR(\xR)=2\log(\mR(\xR)).
    \tag{\PotDefNumber a,b}
\end{align}
 We therefore have expressions for the potential and ionic concentrations over the whole domain and can write the composite asymptotic solutions at leading order via
\begin{subequations}  \label{composites}
    \begin{align}
    \potOG&\sim \pot(x)+\potL(x/\eps)+\potR((1-x)/\eps)-\pot(0)-\pot(1), \label{potComposite}\\
        \pOG &\sim \C(x)+\pL(x/\eps)+\pR((1-x)/\eps)-\C(0)-\C(1),\label{pComposite}\\
        \nOG &\sim \C(x)+\nL(x/\eps)+\nR((1-x)/\eps)-\C(0)-\C(1),\label{nComposite]}
    \end{align}
\end{subequations}
where $\pot$ is defined in \eqref{fullOuterPotential2},  $\C$ is defined in \eqref{Cout2Flux}, $\pL,\nL$ are defined in \hyperref[BoltzmannLeftAnchor]{[\BoltzmannLeftNumber]}, and $\pR,\nR$ are defined by Boltzmann relations for the right inner region,
\begin{align}
\refstepcounter{equation}%
\global\edef\RightConcNumber{\arabic{equation}}% store numeric parent
\phantomsection\label{RightConcAnchor}% create a hyperlink target at this spot
\pR(\xR)&=\frac{(1-\B)^{(1+\rr)}}{n_R^{\rr}}\exp(-\rr\potR(\xR)), \quad \nR(\xR)=n_R\exp(\potR(\xR)).
\tag{\RightConcNumber a,b}
\end{align}

The solutions for the potential ($\pot$, $\potL$ and $\potR$) still have one remaining degree of freedom, $n_R$. This is because the outer leading-order concentrations $\p$ and $\n$ are the same, so the two mass-conservation conditions [\ref{FluxGovEqnsBCs}e,f] represent a duplication of information at leading order. \\

One might expect that we would need to proceed up to $O(\eps)$ in our asymptotic analysis to close this problem. However, this would be overly cumbersome since the closure only requires considering the $O(\eps)$ contributions to the mass-conservation conditions. The mass-conservation conditions [\ref{FluxGovEqnsBCs}e,f] at $O(\eps)$ are
\begin{subequations} \label{Integral Conditions Symm}
\begin{align} 
    0&= \int_{0}^{1} \! p_1(x;n_R)\,  \mathrm{d}x+\int_{0}^{\infty} \! \left(\pL(\xL;n_R)-\C(0)\right)\, \mathrm{d}\xL +\int_{0}^{\infty} \! \left(\pR(\xR;n_R)-\C(1)\right)\, \mathrm{d}\xR, \label{matchCond1}\\
     0&= \int_{0}^{1} \! n_1(x;n_R)\,  \mathrm{d}x+\int_{0}^{\infty} \! \left(\nL(\xL;n_R)-\C(0)\right)\, \mathrm{d}\xL +\int_{0}^{\infty} \! \left(\nR(\xR;n_R)-\C(1)\right)\, \mathrm{d}\xR. \label{matchCond2}
\end{align}
\end{subequations}
Therefore, the only additional contributions that we must determine at $O(\eps)$ in the mass-conservation conditions \eqref{Integral Conditions Symm} are $p_1$ and $n_1$ in the outer region. Determining these contributions will generate an additional degree of freedom; however, since there are two equations in \eqref{Integral Conditions Symm}, there are sufficient constraints to close the problem.\\

To obtain solutions for $p_1,n_1$, we substitute the outer region series expansions \eqref{concDefsSymm} into the Poisson equation \eqref{Poisson} and subsequently find that
\begin{equation}
    p_1(x)=n_1(x):=c_1(x). \label{C1}
\end{equation}
Therefore, the first integral terms in both mass-conservation equations \eqref{Integral Conditions Symm} are equal, and we do not need to solve for $c_1(x)$ explicitly to close the problem: 
we can remove this degree of freedom by rearranging \eqref{Integral Conditions Symm} to obtain
\begin{equation} \label{nReqn}
0
= \int_{0}^{\infty} \!\big(\pL(\xL;n_R)-\nL(\xL;n_R)\big)\,\mathrm{d}\xL
+ \int_{0}^{\infty} \!\big(\pR(\xR;n_R)-\nR(\xR;n_R)\big)\,\mathrm{d}\xR .
\end{equation}
By evaluating the right-hand side of \eqref{nReqn}, we generate an equation for $n_R$ that can be solved to close the problem. \\

In the $\rr=\{1/2, 1, 2\}$ cases, the boundary layer concentrations are expressed in terms of elementary functions and \eqref{nReqn} can be solved analytically.  For all other cases, we solve \eqref{nReqn} numerically in Mathematica. In these cases, the integral terms in \eqref{nReqn} are calculated using \texttt{NIntegrate} for a range of $n_R$ values (i.e. $n_R~\in~[0.05,3.5]$) over a sufficiently large finite domain (i.e. $\xi,\eta=10$) for the integrals to converge. The following simplifications are helpful for calculating the concentration functions from the numerical solutions to \eqref{Gen LBL eqn} and \eqref{Gen RBL Eqn},
\begin{align}
    \refstepcounter{equation}%
\global\edef\SimpleConcNumber{\arabic{equation}}% store numeric parent
\phantomsection\label{SimpleConcAnchor}% create a hyperlink target at this spot
\pL(\xL)&=\frac{(1+\B)^{(1+\rr)}}{(\NN^2 n_R)^{\rr}}\mL(\xL)^{-2 \rr},\quad \nL(\xL)=\NN^2 n_R \mL(\xL)^2,  \tag{\SimpleConcNumber a,b}\\
\quad \pR(\xR)&=\frac{(1-\B)^{(1+\rr)}}{n_R^{\rr}}\mR(\xR)^{-2 \rr}, \quad \nR(\xR)=n_R \mR(\xR)^2.
\tag{\SimpleConcNumber c,d}
\end{align} A fifth-order interpolating function is then constructed for the right-hand side of \eqref{nReqn} as a function of $n_R$ using \texttt{Interpolation}. Finally, \texttt{FindRoot} is used on the interpolating function to determine the value of $n_R$ that satisfies \eqref{nReqn}. Code is available in a GitHub repository\autocite{georgina_ryan_2026_19695204}.

\section*{Results}
The general asymptotic solution of the problem \eqref{FluxGovEqnsInt} with conditions \hyperref[FluxBCsInt]{[\BCsNumber]} for a general \mbox{$z_p\R z_n$} binary electrolyte is given by the composite solutions \eqref{composites} with the value of $n_R$ that solves \eqref{nReqn}. In the $\rr=\{1/2, 1, 2\}$ cases, we can solve the general problem analytically. \\

As an aside, we note that the outer potential solution has a removable singularity when $\B=0$. This corresponds to fluxes for which the ion concentrations are uniform and electromigration alone drives transport. We take the regular limit $\B \to 0$ of  \eqref{fullOuterPotential2} to determine the appropriate expressions when $\B=0$ and $\II>0$:
\begin{align}
    \pot(x)&= V-\II x-\log\left({n_L}\right), \label{OuterPhiL}
\end{align}
where, from \eqref{MultiplyingMatchRight2B}, 
    \begin{align}
        n_L&=n_R \exp\left(V-\II \right). \label{nlnrrelL0}
    \end{align}

\subsection*{Symmetric \texorpdfstring{$z\R z$}{z:z} electrolyte} \label{Sec: z:z eqns} Noting that symmetric electrolytes are well-studied in the literature with alternative boundary conditions\autocite{chenQualitativePropertiesSteadyState1997}, we include the symmetric results for our particular conditions for completeness. For the $\rr=1$ electrolyte, the term inside the square root in \eqref{Gen LBL eqn} simplifies as a difference of squares,
\begin{equation}
    \frac{\mathrm{d}\mL}{\mathrm{d}\xL}= \sqrt{\frac{n_L}{2}}\left(\tl-\mL\right) \left(\tl+\mL\right), \label{zz LBL eqn}
\end{equation}
with boundary and matching condition from \hyperref[mBCsAnchor]{[\mBCsNumber]}. The right boundary layer equation \eqref{Gen RBL Eqn} simplifies identically. 
Solving \eqref{zz LBL eqn} produces a $\tanh$ or $\coth$ solution for $\mL(\xL)$ depending on whether $\tl>1$ or $\tl<1$ (and hence whether we expect a positive or negative gradient at $\mL(0)$). These solutions can be combined using the $\tanh$ addition formula to obtain
\begin{subequations}
    \begin{align} 
             \potL(\xL)&=V+2\log\left( \frac{\tl\left(1+\tl \tanh\left(\tl \NN \sqrt{\dfrac{  n_R}{2}}\xL\right)\right)}{\tl+ \tanh\left(\tl \NN \sqrt{\dfrac{n_R}{2}}\xL\right)}\right), \label{zz left}\\
             \potR(\xR)&=2\log\left( \frac{\tr\left(1+\tr \tanh\left(\tr  \sqrt{\dfrac{ n_R }{2}}\xR\right)\right)}{\tr+ \tanh\left(\tr  \sqrt{\dfrac{ n_R}{2}}\xR\right)}\right). 
            \label{zz right}
    \end{align}
\end{subequations}
Hence, our closure condition for $n_R$ \eqref{nReqn} gives
\begin{align}
 n_R&=\frac{\NN (1-\B )+\B +1}{\NN^2+\NN}. \label{nR Ddef Symm}
\end{align}
From these results, the full composite solutions for the ionic concentrations and electric potential of the \mbox{$z\R z$} electrolyte can be expressed straightforwardly. 
\subsection*{Asymmetric \texorpdfstring{$2 z\R z$}{2z:z} electrolyte}
For the $\rr=2$ electrolyte, the left boundary layer equation  \eqref{Gen LBL eqn} simplifies to 
 \begin{equation}
     \frac{\mathrm{d}\mL}{\mathrm{d}\xL}=\frac{\sqrt{n_L}}{2}(\tl^2-\mL^2) \sqrt{\frac{ \tl^2}{\mL^2}+2}, \label{m2:1eqnL}
 \end{equation} with boundary and matching condition from \hyperref[mBCsAnchor]{[\mBCsNumber]}, and where \eqref{Gen RBL Eqn} simplifies similarly in the right boundary layer. 
 Hence, we find the following results valid for all $\tl,\tr>0$,
\begin{subequations}  \label{r=2}
\begin{align}
    \Phi(\xL)&=V+\log \left(\frac{\tl^2}{2}\right)+\log\left( 3 \left(\frac{\tanh\left( \dfrac{\tl\NN\sqrt{3 n_R}}{2}  \xL\right)+\left(\sqrt{\dfrac{2}{3 \tl ^2}+\dfrac{1}{3}}\right)}{1+\left(\sqrt{\dfrac{2}{3 \tl ^2}+\dfrac{1}{3}}\right)\tanh\left( \dfrac{\tl\NN\sqrt{3 n_R}}{2}  \xL\right)}\right)^2-1\right),
 \label{21right}\\
    \varphi(\xR)&=\log\left(\frac{\tr^2}{2}\right)+\log\left( 3 \left(\frac{\tanh\left( \dfrac{\tr\sqrt{3n_R}}{2}  \xR\right)+\left(\sqrt{\dfrac{2}{3 \tr ^2}+\dfrac{1}{3}}\right)}{1+\left(\sqrt{\dfrac{2}{3 \tr ^2}+\dfrac{1}{3}}\right)\tanh\left( \dfrac{\tr\sqrt{3n_R}}{2}  \xR\right)}\right)^2-1\right), \label{21left}
\end{align}
\end{subequations}
where $n_R$ is defined as the real root of the transcendental equation 
\begin{align}\label{21 nR Eqn}
    &\left(\NN^2 n_R -(1+\B)\right) \sqrt{2 \NN^2 n_R+1+\B }+\left(\NN^2 n_R-\NN^2 (1-\B )\right)  \sqrt{2 n_R+1-\B } =0.
\end{align}

While \eqref{21 nR Eqn} can be inverted in terms of general functions, the resulting expression is not particularly informative and requires tracking different solution branches. As such, we find that a more robust and general approach is to evaluate \eqref{21 nR Eqn} numerically for specific $\NN$ and $\B$ values as required. 

\subsection*{Asymmetric \texorpdfstring{$z\R 2z$}{z:2z} electrolyte} 
In the $\rr=1/2$ case, the left boundary layer equation \eqref{Gen LBL eqn} simplifies to 
 \begin{equation}
     \frac{\mathrm{d}\mL}{\mathrm{d}\xL}=\sqrt{\frac{n_L}{2}}(\tl-\mL) \sqrt{\mL(\mL+2\tl)}, \label{m1:2eqnL}
 \end{equation}  with boundary and matching condition from \hyperref[mBCsAnchor]{[\mBCsNumber]},
 noting that \eqref{Gen RBL Eqn} simplifies similarly in the right boundary layer. 
In a similar manner as the $\rr = 2$ case, we obtain 
\begin{subequations} \label{r=1/2}
\begin{align}
\label{12 left}
  \Phi(\xL)&=V+2\log\left(  \frac{
2\tl \left( \tanh\!\left(\frac{\sqrt{6 n_R}}{4}\NN\tl \xL \right)
+ \sqrt{\frac{3}{1+2 \tl}}\right)^{2}
}{
3\left(1 + \sqrt{\frac{3}{1+2\tl}}
\,\tanh\!\left(\frac{\sqrt{6n_R}}{4}\,\NN \tl\xL \right) \right)^2-
\left(\tanh\!\left(\frac{\sqrt{6n_R}}{4}\,\NN \tl\xL\right)
+ \sqrt{\frac{3}{1+2\tl}}\right)^2
}\right),\\ \label{12right}
    \varphi(\xR)&=2\log\left(  \frac{
2\tr \left( \tanh\!\left(\frac{\sqrt{6 n_R}}{4}\tr \xR \right)
+ \sqrt{\frac{3}{1+2 \tr}}\right)^{2}
}{
3\left(1 + \sqrt{\frac{3}{1+2\tr}}
\,\tanh\!\left(\frac{\sqrt{6 n_R}}{4}\,  \tr \xR\right) \right)^2-
\left(\tanh\!\left(\frac{\sqrt{6n_L}}{4}\,  \tr \xR\right)
+ \sqrt{\frac{3}{1+2\tr}}\right)^2
}\right),
\end{align}
\end{subequations}
where $n_R$ is given by the real root of 
\begin{align}\label{12 nR Eqn}
&\sqrt{3}\Bigl(G(\tr) 
  + G(\tl)\gamma\Bigr) + \Bigl(G(\tr) 
  - 3\gamma(1-\tl)\Bigr)\, H(\tl) 
+ \Bigl(G(\tl)\gamma-3 (1-\tr) \Bigr)\, H(\tr) \notag\\&+ \sqrt{3}\Bigl((-1 + \tr )
  + (-1 + \tl)\gamma\Bigr)\, H(\tl)\,H(\tr)=0,
\end{align}
with
\begin{align}
\refstepcounter{equation}%
\global\edef\HalfSubstitutionNumber{\arabic{equation}}% store numeric parent
\phantomsection\label{HalfSubstitutionAnchor}
    G(\alpha)= (2\alpha + 1)(\alpha - 1), \quad H(\alpha)=\sqrt{1+2\alpha}.
\tag{\HalfSubstitutionNumber a,b}
\end{align}

\begin{comment}
 and $\tl,\tr$ are given by \eqref{alpha Asym def} and  \hyperref[SubRightAnchor]{[\SubRightNumber b]} respectively.   
\end{comment}

\subsection*{Comparison with numerical results}

\begin{figure}
    \centering
        \begin{overpic}[width=\textwidth,tics=5]{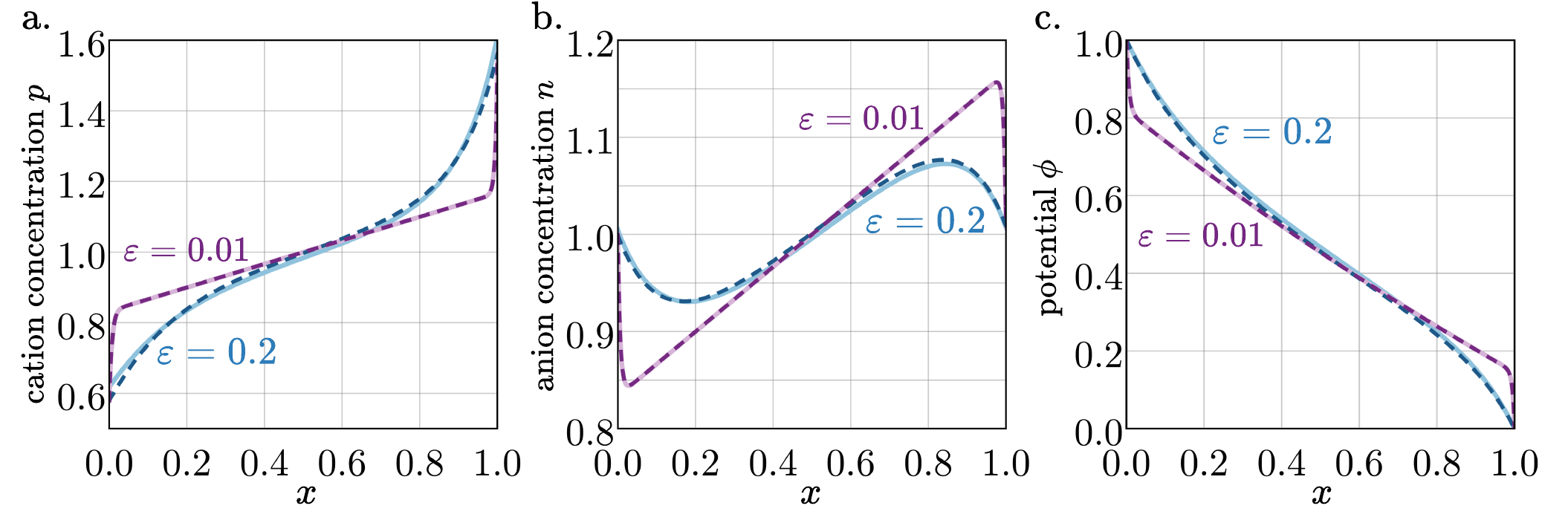}
    \end{overpic}
    \caption{Comparison of numerical (solid) and asymptotic (dashed) solutions for $\rr=2$ with $\eps=0.2$ (blue) and $\eps=0.01$ (purple), with other parameters fixed at $I_p=1, I_n=1,$ and $ V=1.$  As $\eps$ decreases, the asymptotic solution quickly converges to the numerical solution. \textbf{a.} Cation concentration $p$.
\textbf{b.} Anion concentration $n$.
\textbf{c.} Electric potential $\phi$.}
    \label{fig:numerical/asymptotic comparison}
\end{figure}
%Image: Ip=1, In=1, r=2, V=1. 

In \cref{fig:numerical/asymptotic comparison}, we show a comparison of the numerical and asymptotic results for an $\rr=2$ electrolyte to demonstrate their agreement as $\eps$ decreases. When $\eps=0.2$, the overall shape of the concentration curves is qualitatively correct, with a small disagreement between the numerical and asymptotic results. This is most notable in the boundary layers of the cation concentration plot and the outer layer of the anion concentration plot. For $\eps=0.01$, the asymptotic and numerical curves are visually indistinguishable.\\ 
 
Throughout this paper, we generate plots for relatively large values of $\eps$ to see the behaviour within the boundary layers; for realistic values of $\eps$, the boundary layers appear as vertical lines and so the results are difficult to visualise. Importantly, our method is agnostic to the precise value of $\eps$, beyond the requirement that it is small. That is, a key advantage of our asymptotic solutions is that they do not suffer the stiffness issues due to the small parameter $\eps$ that arise in numerical schemes that solve the full PNP equations \autocite{kovalenkoTheoreticalInvestigationPhenomenon2022,flavellConservativeFiniteDifference2014,huFullyDiscretePositivitypreserving2020,chaoIntegralEquationMethod2023}. This is because our asymptotic analysis allows us to systematically and self-consistently scale out the dependence of $\eps$ from the full system to appropriate reduced systems \cref{ImplicitLeft,ImplicitRight} in which $\eps$ does not appear explicitly. Hence, we can straightforwardly generate results for arbitrarily small values of $\eps>0$ without computational difficulty.

%In \cref{fig: 21 comparison analytic v numerical}, we show the same comparison for a $2\R 1$ electrolyte. 

\subsection*{Concentration Boundary Conditions}
Our method allows for a straightforward derivation of solutions when we apply Dirichlet boundary conditions on the concentrations. These boundary conditions are often considered for biological ion channel models \autocite{wangSingularPerturbationSolutions2014, chenQualitativePropertiesSteadyState1997}. The $\rr=1$ solution is given by \citet{chenQualitativePropertiesSteadyState1997}. Below, we provide the explicit solutions for the $\rr=2$ and $\rr=1/2$ cases for concentration boundary conditions (i.e. known boundary concentrations $p_L,n_L, p_R,n_R$ with unknown constant fluxes $I_p,I_n$). We provide a full derivation of these solutions in the Supplementary Material.\\

First, for the $\rr=2$ electrolyte, the outer concentration solution is
\begin{equation}
    c_0(x)=n_L^{2/3} p_L^{1/3}-x \left(n_L^{2/3} p_L^{1/3}-n_R^{2/3} p_R^{1/3}\right),
\end{equation} and the outer solution for the 
potential is 
\begin{equation}
    \pot(x)=V+\frac{1}{3} \log \left(\frac{p_L}{n_L}\right)-\frac{\log \left(\frac{n_R p_L}{n_L p_R}\right)+3 V }{2 \log \left(\frac{n_R}{n_L}\right)+\log \left(\frac{p_R}{p_L}\right)}\log \left(x \left[{\left(\frac{n_R^2 p_R}{n_L^2 p_L}\right)^{1/3}}-1\right]+1\right).
\end{equation}
Then, the left and right boundary layer solutions for the potential are given by \eqref{r=2} with $\tilde{\alpha}_L$ and $\tilde{\alpha}_R$ substituted for $\tl$ and $\tr$ respectively, where
\begin{align}
\refstepcounter{equation}%
\global\edef\ConcAlphasNumber{\arabic{equation}}% store numeric parent
\phantomsection\label{ConcAlphasAnchor}
   \tilde{\alpha}_L=\left(\frac{p_L}{n_L}\right)^{1/6}, \quad \tilde{\alpha}_R=\left(\frac{p_R}{n_R}\right)^{1/6}.
\tag{\ConcAlphasNumber a,b}
\end{align}
These parameters \hyperref[ConcAlphasAnchor]{[\ConcAlphasNumber]} correspond to the limiting parameters from \eqref{alpha Asym def} and \hyperref[SubRightAnchor]{[\SubRightNumber b]} when $\B$ is re-expressed in terms of boundary concentrations instead of prescribed fluxes. \\

%\general{Updated labelling of $\tl$ replacement terms}
Secondly, for the $\rr=1/2$ electrolyte, the outer concentration solution is given by 
\begin{equation}
    c_0(x)=n_L^{1/3} p_L^{2/3}-x \left(n_L^{1/3} p_L^{2/3}-n_R^{1/3} p_R^{2/3}\right),
\end{equation}
and the outer potential solution is
\begin{equation}
    \pot(x)=V+\frac{2}{3} \log \left(\frac{p_L}{n_L}\right)-\frac{2\log \left(\frac{n_R p_L}{n_L p_R}\right)+3 V }{\log \left(\frac{n_R}{n_L}\right)+2\log \left(\frac{p_R}{p_L}\right)}\log \left(x \left[{\left(\frac{n_R p_R^2}{n_L p_L^2}\right)^{1/3}}-1\right]+1\right).
\end{equation}

The boundary layer potential solutions are then given by \eqref{r=1/2} with $\tilde{\alpha}_L$ and $\tilde{\alpha}_R$ substituted for $\tl$ and $\tr$ respectively, where
\begin{equation}
    \tilde{\alpha}_L=\left(\frac{n_L}{p_L}\right)^{1/3}, \quad \tilde{\alpha}_R=\left(\frac{n_R}{p_R}\right)^{1/3}.
\end{equation}
The $\rr= \left\{1/2,2\right\}$ concentration boundary conditions solutions agree with the $\alpha=\rr,\beta=-1$ implicit result from \citet{wangSingularPerturbationSolutions2014}, after accounting for differences in concentration scalings. 

\section*{Discussion}
We now analyse the asymptotic formulae for the general electrolyte and the specific valence ratio cases considered in the previous section. Specifically, we characterise the impact of changing the ion fluxes ($I_p, I_n$) and the valence ratio $\rr$ on the resulting ionic concentrations and electric potential. The behaviour of the potential and ionic concentration curves depends on four parameters: the weighted total current $\II$, the  weighted current imbalance $\B$, the valence ratio $\rr$ and the potential drop $V$. For example, these terms appear in the outer concentration solution \cref{fullOuterPotential2}, and parameters defined in terms of $\B$ and $\rr$ appear in our boundary layer solutions \cref{ImplicitLeft,ImplicitRight}. We recall further that the weighted total current $\II$ and weighted current imbalance $\B$ implicitly depend on the valence ratio $\rr$.

\subsection*{Dependence on ion fluxes}

\begin{figure}[!tb]
    \centering
        \begin{overpic}[width=\textwidth,tics=5]{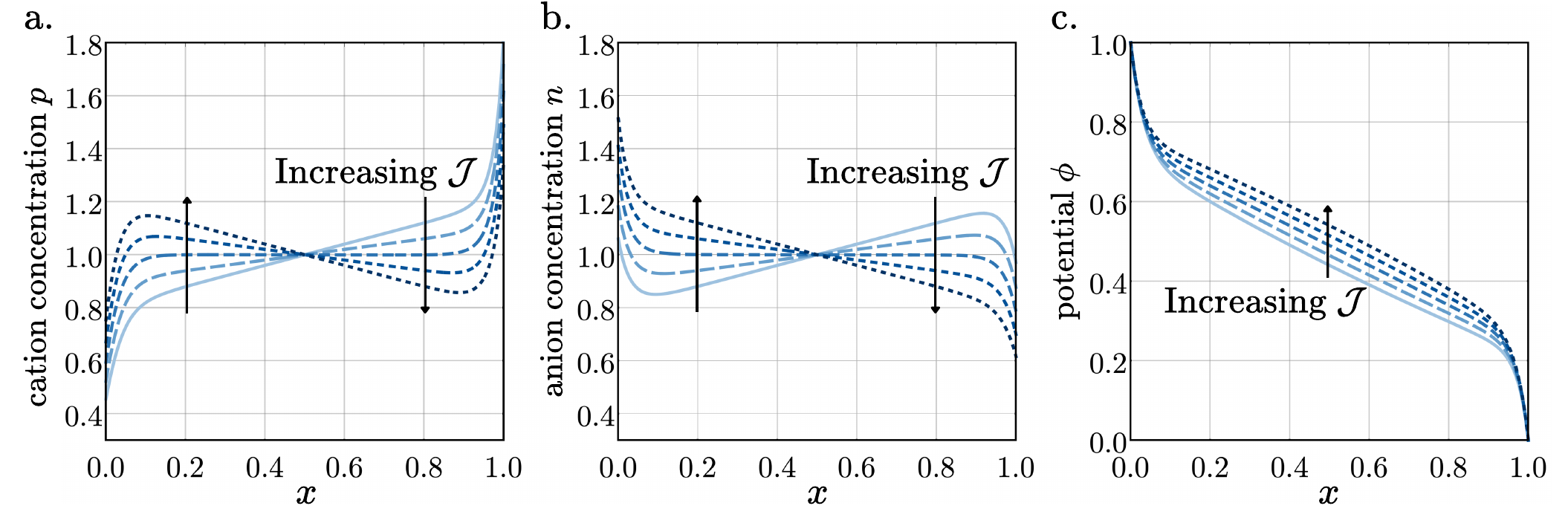}
    \end{overpic}
    \caption{
Concentrations $p,n$ and electric potential $\phi$ for a $2z\!:\!z$ electrolyte
for varying weighted current imbalances $\B=\{-0.2,-0.1,0,0.1,0.2\}$, with fixed $\mathcal{I}=0.5$, $V=1$, and $\varepsilon=0.05$.
Increasing values of $\B$ are distinguished by a monotone
colour gradient from light to dark blue and by progressively finer line dashing.
\textbf{a.} Cation concentration $p$.
\textbf{b.} Anion concentration $n$.
\textbf{c.} Electric potential $\phi$. 
}

    \label{fig:ChangingQ}
\end{figure}
%Figures for Article/Changing Flux Imbalance 21/Plots of Changing Flux Imbalance Q.nb

\begin{figure}[!tb]
    \centering
        \begin{overpic}[width=\textwidth,tics=5]{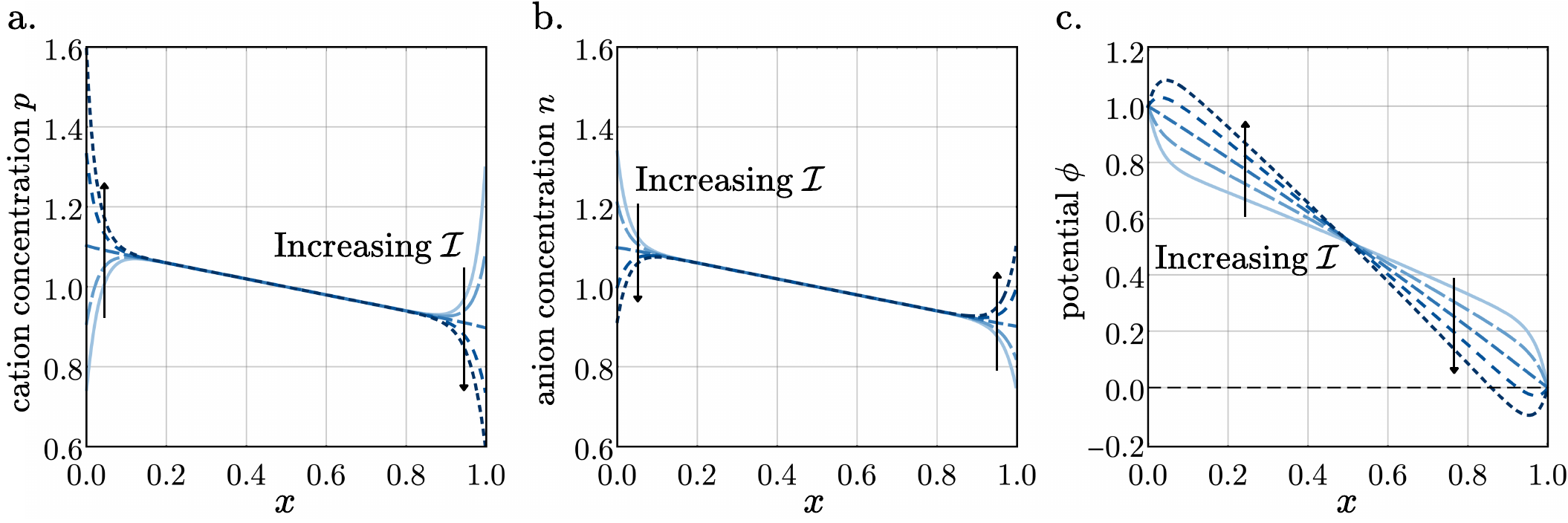}
    \end{overpic}
    \caption{Concentrations $p,n$ and electric potential $\phi$ for a $2z\!:\!z$ electrolyte
for varying total currents $\II=\{0.6,0.8,1.0,1.2,1.4\}$, with fixed $\B=0.1$, $V=1$, and $\varepsilon=0.05$.
Increasing values of $\II$ are distinguished by a monotone
colour gradient from light to dark blue and by progressively finer line dashing.
\textbf{a.} Cation concentration $p$.
\textbf{b.} Anion concentration $n$.
\textbf{c.} Electric potential $\phi$. } 
    \label{fig:ChangingI}
\end{figure}
%Figures for Article/Changing Current 21/Plots of Changing Current I.nb

First, we consider the impact of changing the weighted current imbalance $\B$ when the weighted total current $\II$ is fixed in \cref{fig:ChangingQ}. In \cref{fig:ChangingQ}a,b, we see that $\B$ controls the gradient of the concentrations in the bulk region but has minimal impact on the boundary layers.  When $\B=0$, the concentration in the bulk region is spatially uniform and there is no net movement of particles in the system. When the weighted flux of anions, $rI_n$, is greater than the cation flux $I_p$ (i.e. when $\B<0)$, the gradient is positive and vice versa. Therefore, when $\B>0$, the cations are relatively depleted in the bulk near the cathode ($x=1$) where they are consumed. In \cref{fig:ChangingQ}c, changing the  weighted current imbalance $\B$ has minimal impact on the potential $\phi$ when the weighted total current $\II$ is fixed. The bulk potential increases slightly as $\B$ increases and has a slight curvature change from being convex to concave.\\

% CONCENTRATION AS CURRENT CHANGES
In \cref{fig:ChangingI}, we change the weighted total current $\II=\{0.6,0.8,1.0,1.2,1.4\}$ with a fixed weighted current imbalance $\B$. In \cref{fig:ChangingI}a,b, since the bulk ion concentrations \cref{Cout2Flux} have no $\II$ dependence, only the concentration boundary layers change as $\II$ increases.  For low currents, there is a greater concentration of cations at the cathode and anions at the anode (where they are consumed, respectively), consistent with being in a quasi-equilibrium slightly perturbed from Gouy--Chapman theory \autocite{vansoestbergenDiffusechargeEffectsTransient2010}.  As $\II$ increases, the concentration of cations depletes at the cathode and accumulates at the anode (and vice-versa for the anions). This process of accumulation and depletion in the boundary layers leads to a transition point where there are no visible boundary layers in the concentrations at $\II=1.0$ for this electrolyte. As the weighted total current $\II$ increases further, the accumulation and depletion continues, producing boundary layers where the cations are now depleted at the cathode and accumulate at the anode (and vice-versa for the anions). This behaviour aligns with our physical expectations of observing a limiting current.\\

In \cref{fig:ChangingI}c, we see that the curvature of the potential $\potOG$ initially decreases (becoming more linear) as the weighted total current $\II$ increases. When $\II=1.0$, $\phi$ is linear and there are no boundary layers. As $\II$ increases further, $\phi$ becomes non-monotonic. The gradient of the potential in the bulk region remains negative, so the bulk electric field continues to direct cations to the cathode and anions to the anode. However, the Poisson equation \eqref{Poisson} necessitates that the change in the relative amount of cations and anions corresponds to a change in the curvature of the potential in the diffuse layers. Therefore, the electric field changes direction throughout the cell at high currents. \\

Finally, we make some brief comments on the classical limiting current in this model. From the definitions of the boundary concentrations, such as \hyperref[pLAsymDefAnchor]{[\pLAsymDefNumber a]} and \hyperref[RightPotentialAnchor]{[\RightPotentialNumber a]}, we see that the limiting current condition of zero concentration at the boundary is achieved when $|\B|=1$. This corresponds to logarithmic divergences of the outer potential solution \eqref{fullOuterPotential2}.  The limiting current condition $|\B|=1$ can be expressed equivalently as 
\begin{equation}
    |I_p-\rr I_n|=2(1+\rr). \label{general limiting}
\end{equation}
For the classic example of a symmetric electrolyte with electroactive cations and spectator anions \autocite{levichPhysicochemicalHydrodynamics1962,kontturiIonicTransportProcesses2015}, this becomes \begin{equation}
    |I_p|=4.
\end{equation}
The factor of 4 appears in the classical current scaling to rescale the limiting current for $\rr=1$ electrolytes to unity \autocite{Nernst+1904+52+55, chuElectrochemicalThinFilms2005,bonnefontAnalysisDiffuselayerEffects2001, biesheuvelImposedCurrentsGalvanic2009}. Therefore, \eqref{general limiting} determines the values of the partial current densities $I_p,I_n$ that lead to a limiting current for general binary electrolytes. We note that this model's strict limiting current condition is a restriction on the weighted current imbalance $\B$, not the weighted total current $\II$ (although, the appropriate boundary concentrations will tend to zero in the limit $\II\to \infty$). As an aside, we note that the $O(\eps)$ distinguished limit breaks down near the limiting current and the boundary region widens, allowing super-limiting currents  \autocite{chuElectrochemicalThinFilms2005}, so our observation of the limiting current in this model should be considered solely in the context of providing additional insight into the classical limiting current for asymmetric binary electrolytes. 

%POTENTIAL AS CURRENT CHANGES

%While non-monotonic potentials in electrochemical cells appear in the literature \autocite{biesheuvelImposedCurrentsGalvanic2009,songElectroneutralModelsDynamic2018}, we are not aware of an explicit documentation or discussion of this physical consequence.  

\subsection*{Dependence on ion valence ratio}
\begin{figure}[!tb]
    \centering
    \begin{subfigure}[t]{\textwidth}
        \centering
        \begin{overpic}[width=\textwidth,tics=5]{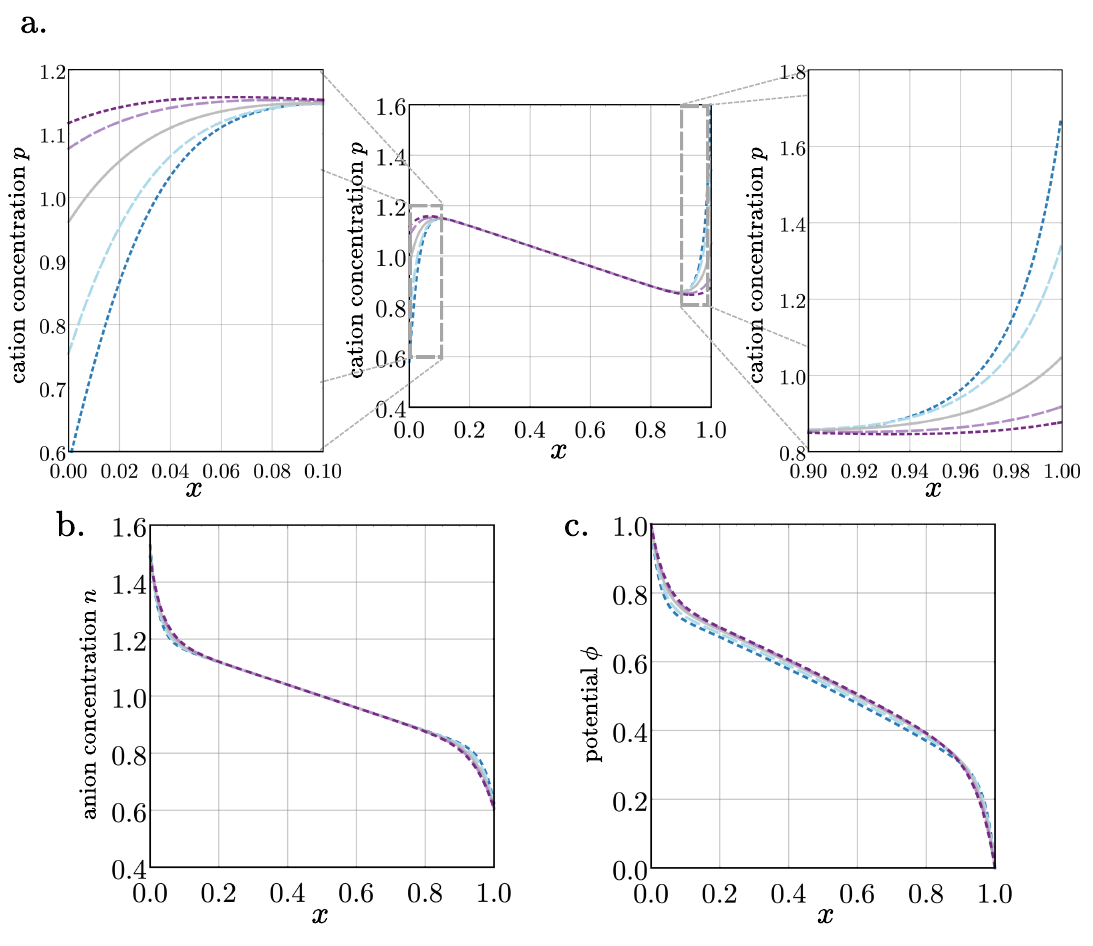}
    \end{overpic}
    \end{subfigure}
    \caption{General electrolytes with $\rr = \{1/3,1/2,1,2,3\}$, transitioning from purple to blue with increasing $\rr$ with $\rr=1$ in grey. We set $\II = 0.5$, $\B = 0.2$, $V = 1$, and $\varepsilon = 0.05$. The same colours are used in each subfigure for the same value of $\rr$. \textbf{a.}~Cation concentration~$p$. \textbf{b.}~Anion concentration~$n$. \textbf{c.}~Electric potential $\potOG$. The cation concentration boundary layers clearly show the impact of changing the valence ratio $\rr$, so we illustrate these with an exploded view. }
    \label{fig:SymmetricZ}
\end{figure}

We now analyse the explicit effect of the ion valence ratio $\rr$ on the electric potential $\potOG$ and the concentrations $\pOG$ and $\nOG$.  In \cref{fig:SymmetricZ}, we consider the ion concentrations and potential for  $\rr = \{1/3, 1/2, 1, 2, 3\}$ with fixed parameters $\II = 0.5$, $\B = 0.2$, $V = 1$, and $\varepsilon = 0.05$. The bulk concentrations in \cref{fig:SymmetricZ}a,b are the same for all $\rr$ values because the outer region concentration \eqref{Cout2Flux} has no explicit $\rr$ dependence. We note, however, that $\B$ is defined in terms of $\rr, I_p,$ and $ I_n$ so these concentration curves have different fluxes ($I_p, I_n$) to balance the varying $\rr$ values. We observe that as the valence ratio $\rr$ increases, the diffuse layers for both ionic concentrations become more extreme (i.e. steeper) according to the boundary layer Boltzmann relations \hyperref[BoltzmannLeftAnchor]{[\BoltzmannLeftNumber]} and \hyperref[RightConcAnchor]{[\RightConcNumber]}. \\

Since our chosen non-dimensionalisation results in the valence ratio $\rr$ factor appearing only in the cation concentration relations \hyperref[BoltzmannLeftAnchor]{[\BoltzmannLeftNumber a]} and \hyperref[RightConcAnchor]{[\RightConcNumber b]}, the boundary layers in \cref{fig:SymmetricZ}a are more extreme than those in \cref{fig:SymmetricZ}b. We therefore additionally use a zoomed-in view to present the cation concentration profile in \cref{fig:SymmetricZ}a.  However, we note that when converted back into dimensional terms and considering the $\B=0$ case (uniform bulk concentrations), the dimensional concentration solutions for a \mbox{$z_p\R z_n$} and \mbox{$z_n\R z_p$} electrolyte are mirror images about $\hat{x}=\hat{L}/2$. Hence, physical interpretation of these plots benefits from converting into dimensional units. \\

Considering the potential $\phi$, in \cref{fig:SymmetricZ}c we see that as the valence ratio $\rr$ increases, the bulk potential translates downwards slightly. This subtle change occurs because the outer region potential \eqref{fullOuterPotential2} has no explicit $\rr$ dependence; the shift is caused solely by the change in the anion boundary concentration $n_L$ and hence depends implicitly on $\rr$ through the closure condition \eqref{nReqn}. The potential boundary layers become steeper as $\rr$ increases, resulting in larger differences between the cation and anion concentrations ($p,n$) in the boundaries through the Poisson equation \eqref{Poisson}.  \\

%$\II = 0.5$, $\B = 0.2$, $V = 1$, and $\varepsilon = 0.05$.

From our asymptotic results, we can derive an analytic formula for the bulk potential value $\critpot$ for a general binary electrolyte in the no-flux/blocking-electrode case (\mbox{$I_p=I_n=0$}). In this case, the right-hand side of \eqref{PotentialDE} is zero. Therefore, the outer potential is an unknown constant $\critpot$ that must be determined by matching into the boundary layers. The constant outer potential solution $\critpot$ could be used as an effective boundary condition for situations where only information about the bulk potential is required with blocking electrodes. 
We integrate the Poisson equation \eqref{Poisson} over the domain and apply the mass-conservation conditions to obtain 
\begin{align}
\eps^2\int_{0}^{1} \!    \frac{\mathrm{d}^2\phi}{\mathrm{d}x^2}\, \mathrm{d}x&=-\int_{0}^{1} \! \pOG \, \mathrm{d}x +\int_{0}^{1} \!  \nOG \,\mathrm{d}x=0.   \label{nofluxMassEqn}
\end{align}
Transforming into our boundary layer variables $\xL,\xR$, \eqref{nofluxMassEqn} is equivalently written as
\begin{align}
\frac{\mathrm{d}\potL}{\mathrm{d}\xL}|_{{\xL=0}}&=-\frac{\mathrm{d}\potR}{\mathrm{d}\xR}|_{{\xR=0}}.\label{conditionsMassCons}
\end{align}
%Therefore, the potential gradient at boundaries is directly connected to the mass-conservation condition. 
Note that \eqref{conditionsMassCons} applies for all values of $I_p,I_n$ and is automatically satisfied by our boundary layer solutions \cref{Gen LBL eqn,Gen RBL Eqn}.  In the no-flux case, \eqref{conditionsMassCons} closes the problem and gives
\begin{align}
    \critpot(r,V)=\frac{1}{1+\rr}\log \left(\frac{\rr \left(\ee{ V}-1\right)}{1-\ee{-\rr V}} \right) . \label{CriticalPotentialEqn}
\end{align}
Therefore, the bulk potential value $\critpot$ in the blocking-electrodes case is determined by the valence ratio $\rr$ and the potential drop $V$. We note that \eqref{OuterPhiL} is equivalent to \eqref{CriticalPotentialEqn} when $\II=\B=0$.
The impacts of valence on this bulk potential \eqref{CriticalPotentialEqn} are most easily interpreted in dimensional quantities. Converting this bulk potential back to its dimensional form $\hat{\phi}_0$, we find 
\begin{align}
    \lim_{z_n \to \infty} \hat{\phi}_0&=\lim_{z_n \to \infty} \frac{RT}{Fz_n} \critpot\left(\frac{z_p}{z_n}, \frac{F z_n}{RT} \hat{V}\right)=\hat{V},\\
    \lim_{z_p \to \infty} \hat{\phi}_0&=\lim_{z_p \to \infty} \frac{RT}{Fz_n} \critpot\left(\frac{z_p}{z_n}, \frac{F z_n}{RT} \hat{V}\right)=0.
\end{align}
Therefore, as the magnitude of the anion valence increases, the dimensional outer potential increases, and vice versa for the cation valence.

\subsection*{Characterising the intermediate-current transition}
\begin{figure}
    \centering
    \includegraphics[width=0.8\textwidth]{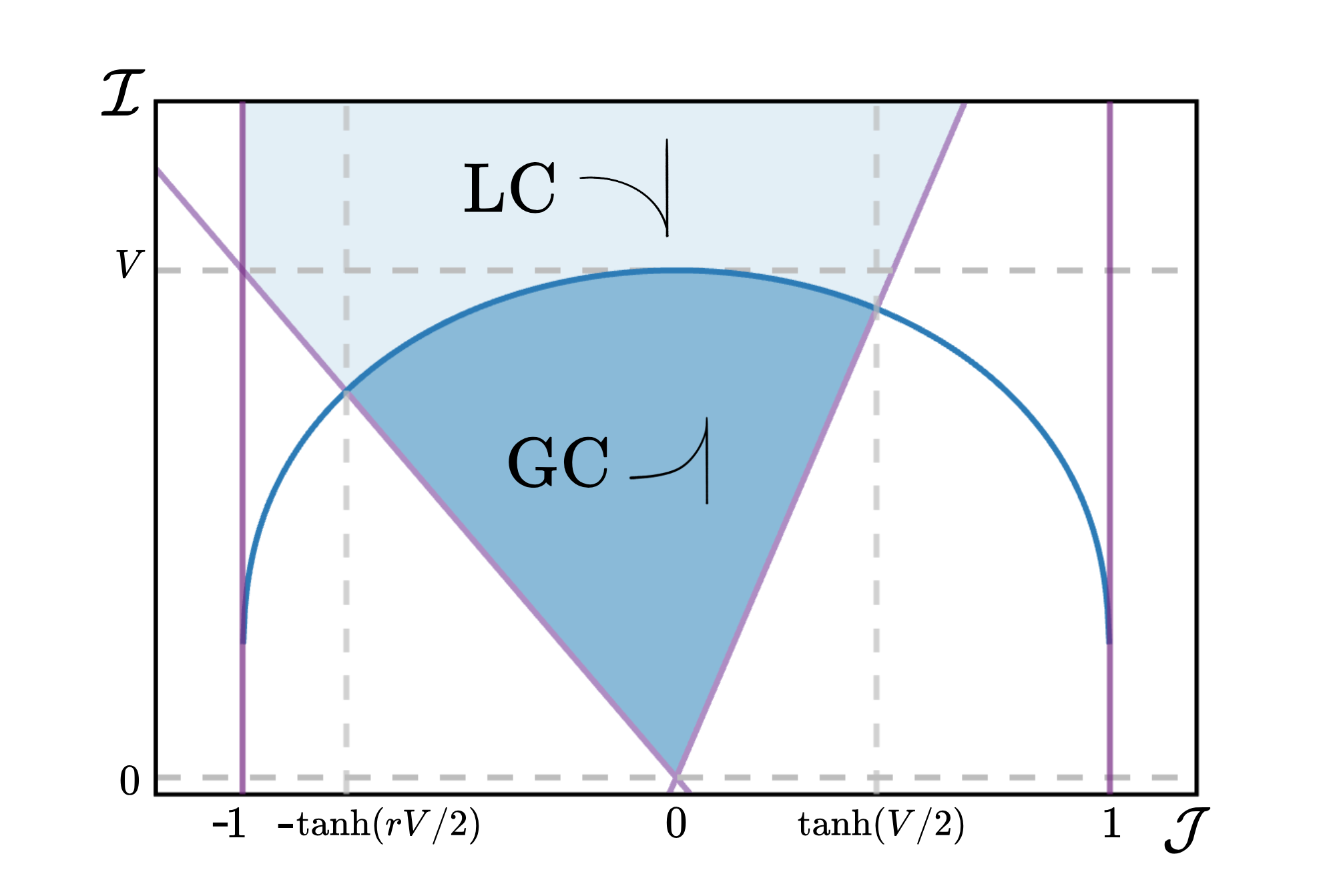}
    \caption{Phase diagram of the weighted total current $\II$ versus the weighted current imbalance $\B$ showing the transition in qualitative behaviour of the system. Region GC in dark blue is the region of $\II$--$\B$ space for a general binary electrolyte that has cations accumulating at the cathode (near-Gouy--Chapman regime). Region LC in light blue has cations depleted at the cathode (the near-limiting current regime). The dark blue line is \eqref{I limit Gen}, and the purple lines mark the physically relevant domain given by \eqref{L Limit}. We indicate the visual asymmetry of these representative regions for an $\rr>1$ electrolyte. }
    \label{fig: Pizza Slice General}
\end{figure}

From our asymptotic analysis, we have shown that the qualitative behaviour of the ion concentrations and electric potential depend on the weighted total current $\II$, the weighted current imbalance $\B$, and the valence ratio $\rr$. We now use the asymptotic results to characterise the transition behaviour seen in our numerical simulations by collapsing our results onto a phase diagram of $\II$ versus $\B$ in \cref{fig: Pizza Slice General}.\\

To create the phase diagram, we first divide the behaviour we have observed in our numerical results (see \cref{fig:numerical initial results}) into two distinct qualitative regimes: the near-Gouy--Chapman regime (GC), where cations accumulate at the cathode relative to the bulk, and the limiting-current regime (LC), where cations deplete at the cathode relative to the bulk. The observed transition where boundary layers disappear separates these two regions. \\

The physically relevant values for $\B$ to consider for the phase diagram correspond to $I_p,I_n\ge0$ (fluxes at the boundaries in the expected directions) and $|\B|<1$ (below limiting current). Therefore, we impose the following physical restriction on $\B$,
\begin{equation}
-\min\!\left(1,\frac{\II r}{2}\right)
\;<\;
\B
\;<\;
\min\!\left(1,\frac{\II}{2}\right)
. \label{L Limit}
\end{equation}
When $\B=0$, the system has a uniform unit bulk concentration (see \eqref{Cout2Flux}) and therefore the transition point occurs for $p_L=n_L=1$ and $p_R=n_R=1$. Using the relationship between $n_R$ and $n_L$ in the $\B=0$ case \eqref{nlnrrelL0}, we immediately see that the transition point for a general electrolyte when $\B=0$ occurs at 
\begin{equation}
    \II= V. \label{MaxTransCurrent}
\end{equation}

Next, we derive an explicit expression for the transition line between Region GC and Region LC. From the definition of $p_L$ in terms of $n_L$ \hyperref[pLAsymDefAnchor]{[\pLAsymDefNumber a]}, we find that the transition where $p_L=n_L$ occurs when $n_L=1+\B$. Likewise on the right side, $p_R=n_R$ when $n_R=1-\B$. Therefore, substituting these critical values of $n_L,n_R$ into the expression for $n_L$ in terms of $n_R$ \eqref{MultiplyingMatchRight2B}, we find the transition between Region GC and Region LC occurs at a critical current,
%\begin{equation}
  %  \bar{\II}=\frac{2(1+\rr)\B V}{\log(2(1+\rr)+2(1+\rr)\B)-\log(2(1+\rr)-2(1+\rr)\B)}.\label{I limit Gen}
%\end{equation}
\begin{equation}
    \bar{\II}=\frac{2\B V}{\log(1+\B)-\log(1-\B)}.\label{I limit Gen}
\end{equation}
The range of physically relevant $\B$ values on the transition line is \begin{equation}
   \B\in[ -\tanh\left(\rr V/2\right),\tanh\left(V/2\right)].
\end{equation}
The extremal transition values of $\B$ tend to $\pm 1$ in the limit $V\to \infty$. Therefore, ionic concentrations in Region GC can tend to zero in the limit of infinite voltage. However, in our minimal model (ignoring the change in distinguished limit near the limiting current \autocite{chuElectrochemicalThinFilms2005}), a limiting current is achievable for systems in Region LC with a finite voltage.

\section*{Conclusion}
We have considered a one-dimensional steady-state Poisson--Nernst--Planck (PNP) minimal model of an electrolytic cell with constant-flux boundary conditions for a general binary electrolyte with valence ratio $\rr=z_p/z_n$, where $z_p$ and $z_n$ are the cation and anion valences, respectively. The constant-flux boundary conditions represent Faradaic reactions at the electrodes and hence a current throughout the cell (denoted by $I_p$ for the cation concentration and $I_n$ for the anion concentration), separating this work from studies of valence asymmetry with the Poisson-Boltzmann equations. From an initial numerical study of our minimal PNP equations, we identified a continuous, smooth transition from agreeing with Gouy--Chapman theory at zero current to limiting-current behaviour at high currents. We identified a transition point at a critical intermediate current that separates two distinct qualitative regimes (one where cations accumulate at the cathode relative to the electroneutral bulk concentration, and another where they are depleted). At this critical current, the potential has a linear profile, so the electric field throughout the cell is constant, and the electrolyte is electroneutral over the whole domain. We found that the valence ratio $\rr$ controls this transition.\\

In order to characterise how the valence ratio $\rr$ impacts the system, we reduced our minimal model using asymptotic analysis. This approach was possible because of the smallness of the parameter $\eps$, which represents the ratio of the characteristic length of the diffuse layer near an electrode to the length of the cell, in the dimensionless PNP equations. We derived solutions for the cation concentration $p$, anion concentration $n$, and electric potential $\phi$ in three distinct regions: the electroneutral bulk and two diffuse layers (boundary layers), one near each electrode. Through our analysis, we have characterised how this minimal model depends on the valence ratio $\rr$ explicitly and also implicitly through two dimensionless parameters $\mathcal{I}$ and $\mathcal{J}$, which represent a dimensionless weighted total current and dimensionless weighted current imbalance, respectively, and which contain the cation and anion fluxes.\\

For our model, we produced explicit analytic expressions for the composite asymptotic solutions for the $\rr=\{1/2, 1, 2\}$  cases and we additionally generated implicit solutions for asymmetric electrolytes for any value of $\rr$. Our implicit solutions are simple to evaluate numerically for any value of $\eps$, as our asymptotic techniques allow us to systematically scale out this small parameter from the appropriate reduced governing differential equations we derive. This overcomes a key difficulty faced by numerical modelling of stiff PNP systems. With slight modifications provided in the Supplementary Material, these solutions can also be used with Dirichlet concentration conditions, which are relevant to many mathematical biology applications, such as ion channels in cell membranes. While implicit asymptotic solutions for concentration boundary conditions for general electrolytes are known\autocite{wangSingularPerturbationSolutions2014}, explicit elementary function representations of the $\rr=\{1/2,2\}$ solutions are valuable as they require no additional numerical calculations. \\

We produced a phase diagram (\cref{fig: Pizza Slice General})  that identifies whether the system will display the behaviour of cations accumulating or depleting at the cathode relative to the bulk (signifiers of the two distinct qualitative regimes) in terms of the valence ratio $\rr$ and the cation and anion concentration fluxes $I_p, I_n$. This phase diagram explains how binary electrolytes with different valence ratios can show significantly different qualitative behaviour for identical cation and anion fluxes, explaining observations that we had made from initial numerical simulations. \\

A natural extension to this work would be to incorporate Butler--Volmer reaction conditions\autocite{newmanElectrochemicalSystems2021,richardsonTimedependentModellingAsymptotic2007} to explore how this valence ratio-dependent transition behaviour persists once interfacial reaction kinetics can limit the transport-driven effects. Additionally, we could extend the time-evolution work of \citet{jarveyIonTransportElectrochemical2022} to explore how the system evolves temporally to the steady-states that we have found for different valence ratios. The model in \citet{jarveyIonTransportElectrochemical2022} also captures the Stern layer, which we could incorporate into our analysis. \\

Our model analysis has shown that the valence ratio of a binary electrolyte determines many key qualities of the steady-state ionic concentrations and electric potential, including how the system transitions from the Gouy--Chapman regime towards limiting current behaviour as the total current increases. Our results provide a comprehensive picture of how multivalency in a binary electrolyte can play a fundamental role in the steady state of non-equilibrium electrochemical systems.  

\section*{Acknowledgements}
G. C. Ryan acknowledges funding from the Commonwealth Bank John Monash Scholarship (General Sir John Monash Foundation, Australia).

\section*{Copyright}
For the purpose of open access, the author has applied a CC BY public copyright licence to any author accepted manuscript arising from this submission.
\label{Bibliography}

%\bibliographystyle{siam}  % Use the "unsrtnat" BibTeX style for formatting the Bibliography
%\bibliography{Bibliography}  % The references (bibliography) information are stored in the file named "Bibliography.bib"
\sloppy

\printbibliography

@article{chenQualitativePropertiesSteadyState1997,
  title = {Qualitative {{properties}} of {{steady-state Poisson--Nernst--Planck systems}}: {{perturbation}} and {{simulation study}}},
  shorttitle = {Qualitative {{Properties}} of {{Steady-State Poisson--Nernst--Planck Systems}}},
  author = {Barcilon, V. and Chen, D. P. and Eisenberg, R. S. and Jerome, J. W.},
  year = {1997},
  month = jun,
  journal = {SIAM J. Appl. Math.},
  volume = {57},
  number = {3},
  pages = {631--648},
  issn = {0036-1399, 1095-712X},
  doi = {10.1137/S0036139995312149},
  urldate = {2025-03-21},
  langid = {english}
}

@book{bardElectrochemicalMethodsFundamentals2022,
  title = {Electrochemical Methods: Fundamentals and Applications},
  shorttitle = {Electrochemical Methods},
  author = {Bard, Allen J. and Faulkner, Larry R. and White, Henry S.},
  date = {2022},
  edition = {Third edition},
  publisher = {Wiley},
  location  = {Hoboken, NJ},
  isbn = {978-1-119-33406-4},
  langid = {english},
  pagetotal = {1044}
}

@article{chuElectrochemicalThinFilms2005,
  title = {Electrochemical {{thin films}} at and above the {{classical limiting current}}},
  author = {Chu, Kevin T. and Bazant, Martin Z.},
  year = {2005},
  month = jan,
  journal = {SIAM J. Appl. Math.},
  volume = {65},
  number = {5},
  pages = {1485--1505},
  issn = {0036-1399, 1095-712X},
  doi = {10.1137/040609926},
  urldate = {2025-03-21},
  langid = {english}
}

@article{bazantCurrentVoltageRelationsElectrochemical2005,
  title = {Current-{{voltage relations}} for {{electrochemical thin films}}},
  author = {Bazant, Martin Z. and Chu, Kevin T. and Bayly, B. J.},
  date = {2005-01},
  journaltitle = {SIAM J. Appl. Math.},
  shortjournal = {SIAM J. Appl. Math.},
  volume = {65},
  number = {5},
  pages = {1463--1484},
  issn = {0036-1399, 1095-712X},
  doi = {10.1137/040609938},
  url = {http://epubs.siam.org/doi/10.1137/040609938},
  urldate = {2025-05-19},
  langid = {english}
}

@article{jarveyIonTransportElectrochemical2022,
  title = {Ion {{transport}} in an {{electrochemical cell}}: {{a theoretical framework}} to {{couple dynamics}} of {{double layers}} and {{redox reactions}} for {{multicomponent electrolyte solutions}}},
  shorttitle = {Ion {{Transport}} in an {{Electrochemical Cell}}},
  author = {Jarvey, Nathan and Henrique, Filipe and Gupta, Ankur},
  date = {2022-09-01},
  journaltitle = {J. Electrochem. Soc.},
  shortjournal = {J. Electrochem. Soc.},
  volume = {169},
  number = {9},
  pages = {093506},
  issn = {0013-4651, 1945-7111},
  doi = {10.1149/1945-7111/ac908e},
  url = {https://iopscience.iop.org/article/10.1149/1945-7111/ac908e},
  urldate = {2025-06-05},
  langid = {english}
}

@article{chapmanLIContributionTheory1913,
  title = {{{LI}}. {{A}} contribution to the theory of electrocapillarity},
  author = {Chapman, David Leonard},
  date = {1913-04},
  journaltitle = {Lond. Edinb. Dublin Philos. Mag. J. Sci.},
  shortjournal = {The London, Edinburgh, and Dublin Philosophical Magazine and Journal of Science},
  volume = {25},
  number = {148},
  pages = {475--481},
  issn = {1941-5982, 1941-5990},
  doi = {10.1080/14786440408634187},
  url = {https://www.tandfonline.com/doi/full/10.1080/14786440408634187},
  urldate = {2025-05-19},
  langid = {english}
}

@book{kontturiIonicTransportProcesses2015,
	address = {Oxford},
	title = {Ionic {Transport} {Processes} in {Electrochemistry} and {Membrane} {Science}},
	isbn = {978-0-19-871999-1},
	publisher = {Oxford University Press},
	author = {Kontturi, Kyosti and Murtom\"{a}ki, Lasse and Manzanares, Jose A.},
	year = {2015},
}

@article{gouyConstitutionChargeElectrique1910,
	title = {Sur la constitution de la charge électrique à la surface d'un électrolyte},
	volume = {9},
	issn = {0368-3893},
	url = {http://www.edpsciences.org/10.1051/jphystap:019100090045700},
	doi = {10.1051/jphystap:019100090045700},
	number = {1},
	urldate = {2025-05-19},
	journal = {J. Phys. (Paris)},
	author = {Gouy, M.},
	year = {1910},
	pages = {457--468},
}

@article{bazantDiffusechargeDynamicsElectrochemical2004,
	title = {Diffuse-charge dynamics in electrochemical systems},
	volume = {70},
	copyright = {http://link.aps.org/licenses/aps-default-license},
	issn = {1539-3755, 1550-2376},
	url = {https://link.aps.org/doi/10.1103/PhysRevE.70.021506},
	doi = {10.1103/PhysRevE.70.021506},
	language = {en},
	number = {2},
	urldate = {2025-03-21},
	journal = {Phys. Rev. E},
	author = {Bazant, Martin Z. and Thornton, Katsuyo and Ajdari, Armand},
	month = aug,
	year = {2004},
	pages = {021506},
}

@article{andriettiExactSolutionUnidimensional1976,
	title = {Exact solution of the unidimensional {Poisson}--{Boltzmann} equation for a 1:2 (2:1) electrolyte},
	volume = {16},
	copyright = {https://www.elsevier.com/tdm/userlicense/1.0/},
	issn = {00063495},
	shorttitle = {Exact solution of the unidimensional {Poisson}-{Boltzmann} equation for a 1},
	url = {https://linkinghub.elsevier.com/retrieve/pii/S000634957685761X},
	doi = {10.1016/S0006-3495(76)85761-X},
	number = {9},
	urldate = {2025-08-12},
	journal = {Biophys. J.},
	author = {Andrietti, F. and Peres, A. and Pezzotta, R.},
	month = sep,
	year = {1976},
	pages = {1121--1124},
}

@book{newmanElectrochemicalSystems2021,
	address = {Hoboken, NJ},
	edition = {4},
	title = {Electrochemical {Systems}},
	isbn = {978-1-119-51460-2},
	publisher = {John Wiley \& Sons Inc.},
	author = {Newman, John and Balsara, Nitash P.},
	year = {2021},
}

@book{levichPhysicochemicalHydrodynamics1962,
	address = {Englewood Cliffs, NJ},
	edition = {1},
	series = {Prentice-{Hall} {International} {Series} in the {Physical} and {Chemical} {Engineering} {Sciences}},
	title = {Physicochemical {Hydrodynamics}},
	publisher = {Prentice-Hall},
	author = {Levich, Veniamin},
	year = {1962},
}

@article{wangSingularPerturbationSolutions2014,
	title = {Singular perturbation solutions of steady-state {Poisson}--{Nernst}--{Planck} systems},
	volume = {89},
	copyright = {http://link.aps.org/licenses/aps-default-license},
	issn = {1539-3755, 1550-2376},
	url = {https://link.aps.org/doi/10.1103/PhysRevE.89.022722},
	doi = {10.1103/PhysRevE.89.022722},
	number = {2},
	urldate = {2025-03-21},
	journal = {Phys. Rev. E},
	author = {Wang, Xiang-Sheng and He, Dongdong and Wylie, Jonathan J. and Huang, Huaxiong},
	month = feb,
	year = {2014},
	pages = {022722}
}

@article{biesheuvelImposedCurrentsGalvanic2009,
  title = {Imposed currents in galvanic cells},
  author = {Biesheuvel, P.M. and van Soestbergen, M. and Bazant, M.Z.},
  date = {2009-08},
  journaltitle = {Electrochim. Acta},
  shortjournal = {Electrochimica Acta},
  volume = {54},
  number = {21},
  pages = {4857--4871},
  issn = {00134686},
  doi = {10.1016/j.electacta.2009.03.073},
  url = {https://linkinghub.elsevier.com/retrieve/pii/S0013468609004915},
  urldate = {2025-09-09},
  langid = {english}
}

@article{huFullyDiscretePositivitypreserving2020,
  title = {A fully discrete positivity-preserving and energy-dissipative finite difference scheme for Poisson--Nernst--{{Planck}} equations},
  author = {Hu, Jingwei and Huang, Xiaodong},
  date = {2020-05},
  journaltitle = {Numer. Math.},
  shortjournal = {Numer. Math.},
  volume = {145},
  number = {1},
  pages = {77--115},
  issn = {0029-599X, 0945-3245},
  doi = {10.1007/s00211-020-01109-z},
  url = {http://link.springer.com/10.1007/s00211-020-01109-z},
  urldate = {2025-05-19},
  langid = {english}
}

@article{flavellConservativeFiniteDifference2014,
  title = {A conservative finite difference scheme for {{Poisson}}--{{Nernst}}--{{Planck}} equations},
  author = {Flavell, Allen and Machen, Michael and Eisenberg, Bob and Kabre, Julienne and Liu, Chun and Li, Xiaofan},
  date = {2014-03},
  journaltitle = {J. Comput. Electron.},
  shortjournal = {J Comput Electron},
  volume = {13},
  number = {1},
  pages = {235--249},
  issn = {1569-8025, 1572-8137},
  doi = {10.1007/s10825-013-0506-3},
  url = {http://link.springer.com/10.1007/s10825-013-0506-3},
  urldate = {2025-05-19},
  langid = {english}
}

@article{beshaDesignMonovalentIon2019,
  title = {Design of monovalent ion selective membranes for reducing the impacts of multivalent ions in reverse electrodialysis},
  author = {Besha, Abreham Tesfaye and Tsehaye, Misgina Tilahun and Aili, David and Zhang, Wenjuan and Tufa, Ramato Ashu},
  year = 2019,
  month = dec,
  journal = {Membranes},
  volume = {10},
  number = {1},
  pages = {7},
  issn = {2077-0375},
  doi = {10.3390/membranes10010007},
  urldate = {2026-02-04},
  langid = {english}
}

@article{guduruBriefReviewMultivalent2016,
  title = {A brief review on multivalent intercalation batteries with aqueous electrolytes.},
  author = {Guduru, Ramesh and Icaza, Juan},
  year = 2016,
  month = feb,
  journal = {Nanomaterials},
  volume = {6},
  number = {3},
  pages = {41},
  issn = {2079-4991},
  doi = {10.3390/nano6030041},
  urldate = {2026-02-04},
  langid = {english}
}

@article{postInfluenceMultivalentIons2009,
  title = {Influence of multivalent ions on power production from mixing salt and fresh water with a reverse electrodialysis system},
  author = {Post, Jan W. and Hamelers, Hubertus V.M. and Buisman, Cees J.N.},
  year = 2009,
  month = mar,
  journal = {J. Membr. Sci.},
  volume = {330},
  number = {1-2},
  pages = {65--72},
  issn = {03767388},
  doi = {10.1016/j.memsci.2008.12.042},
  urldate = {2026-01-28},
  copyright = {https://www.elsevier.com/tdm/userlicense/1.0/},
  langid = {english}
}

@article{shahReviewRecentAdvances2025,
  title = {Review of recent advances in multivalent ion batteries for next generation energy storage},
  author = {Shah, Raj and Marussich, Kate and Mittal, Vikram},
  year = 2025,
  month = dec,
  journal = {Electrochem}, %note: this isn't an abbreviation! The full title is electrochem
  volume = {6},
  number = {4},
  pages = {44},
  issn = {2673-3293},
  doi = {10.3390/electrochem6040044},
  urldate = {2026-02-04},
  abstract = {As demand for high-performance energy storage grows across grid and mobility sectors, multivalent ion batteries (MVIBs) have emerged as promising alternatives to lithium-based systems due to their potential for higher volumetric energy density and material abundance. This review comprehensively examines recent breakthroughs in magnesium, zinc, aluminum, and calcium-based battery chemistries, with a focus on overcoming barriers related to slow ion transport, limited reversibility, and electrode degradation. Advances in aqueous and non-aqueous electrolyte formulations, including solvation shell engineering, interfacial passivation, and dual-zone ion transport, are discussed for their role in improving compatibility and cycling stability. Particular focus is placed on three high-impact innovations: solvation-optimized Mg-ion systems for improved mobility and retention, interface-engineered Zn-ion batteries enabling dendrite-free operation, and sustainable Al-ion technologies targeting grid-scale deployment with eco-friendly electrolytes and recyclable materials. Cross-cutting insights from operando characterization techniques and AI-guided materials discovery are also evaluated for their role in accelerating MVIB development. By integrating fundamental materials innovation with practical system design, multivalent ion batteries offer a compelling path toward next-generation, safer, and more sustainable energy storage platforms.},
  langid = {english},
  file = {/Users/georginaryan/Desktop/DPhil/Research/Draft Papers/Journal Article - Chapter 4/Comparison Articles/Introduction context articles/electrochem-06-00044-v2.pdf}
}

@article{wuMitigatingInfluenceMultivalent2024,
  title = {Mitigating the influence of multivalent ions on power density performance in a single-membrane capacitive reverse electrodialysis cell},
  author = {Wu, Nan and Levant, Michael and Brahmi, Youcef and Tregouet, Corentin and Colin, Annie},
  year = 2024,
  month = jul,
  journal = {Sci. Rep.},
  volume = {14},
  number = {1},
  pages = {16984},
  issn = {2045-2322},
  doi = {10.1038/s41598-024-67690-7},
  urldate = {2026-02-04},
  langid = {english}
}

@article{bonnefontAnalysisDiffuselayerEffects2001,
  title = {Analysis of diffuse-layer effects on time-dependent interfacial kinetics},
  author = {Bonnefont, Antoine and Argoul, Fran{\c c}oise and Bazant, Martin Z.},
  year = 2001,
  month = mar,
  journal = {J. Electroanal. Chem.},
  volume = {500},
  number = {1-2},
  pages = {52--61},
  issn = {15726657},
  doi = {10.1016/S0022-0728(00)00470-8},
  urldate = {2025-05-19},
  copyright = {https://www.elsevier.com/tdm/userlicense/1.0/},
  langid = {english}
}

@article{guptaElectricalDoubleLayers2018,
  title = {Electrical double layers: effects of asymmetry in electrolyte valence on steric effects, dielectric decrement, and ion–ion correlations},
  shorttitle = {Electrical {{Double Layers}}},
  author = {Gupta, Ankur and Stone, Howard A.},
  date = {2018-10-09},
  journaltitle = {Langmuir},
  shortjournal = {Langmuir},
  volume = {34},
  number = {40},
  pages = {11971--11985},
  issn = {0743-7463, 1520-5827},
  doi = {10.1021/acs.langmuir.8b02064},
  url = {https://pubs.acs.org/doi/10.1021/acs.langmuir.8b02064},
  urldate = {2026-02-05},
  langid = {english}
}

@article{xingPoissonBoltzmannTheoryTwo2011,
  title = {Poisson-{{Boltzmann}} theory for two parallel uniformly charged plates},
  author = {Xing, Xiangjun},
  date = {2011-04-27},
  journaltitle = {Phys. Rev. E},
  shortjournal = {Phys. Rev. E},
  volume = {83},
  number = {4},
  pages = {041410},
  issn = {1539-3755, 1550-2376},
  doi = {10.1103/PhysRevE.83.041410},
  url = {https://link.aps.org/doi/10.1103/PhysRevE.83.041410},
  urldate = {2026-04-24},
  langid = {english}
}

@article{zhangExactSolutionNonlinear2018,
  title = {An exact solution of the nonlinear {{Poisson--Boltzmann}} equation in parallel-plate geometry},
  author = {Zhang, Wenyao and Wang, Qiuwang and Zeng, Min and Zhao, Cunlu},
  date = {2018-11},
  journaltitle = {Colloid Polym. Sci.},
  shortjournal = {Colloid Polym Sci},
  volume = {296},
  number = {11},
  pages = {1917--1923},
  issn = {0303-402X, 1435-1536},
  doi = {10.1007/s00396-018-4394-8},
  url = {http://link.springer.com/10.1007/s00396-018-4394-8},
  urldate = {2026-02-18},
  langid = {english}
}

@article{richardsonTimedependentModellingAsymptotic2007,
  title = {Time-dependent modelling and asymptotic analysis of electrochemical cells},
  author = {Richardson, G. and King, J. R.},
  date = {2007-10-15},
  journaltitle = {J. Eng. Math.},
  shortjournal = {J Eng Math},
  volume = {59},
  number = {3},
  pages = {239--275},
  issn = {0022-0833, 1573-2703},
  doi = {10.1007/s10665-006-9114-6},
  url = {http://link.springer.com/10.1007/s10665-006-9114-6},
  urldate = {2025-05-19},
  langid = {english}
}

@book{byrdHandbookEllipticIntegrals1971,
  title     = {Handbook of Elliptic Integrals for Engineers and Scientists},
  author    = {Byrd, Paul F. and Friedman, Morris D.},
  year      = {1971},
  publisher = {Springer},
  address   = {Berlin}
}

@software{Mathematica,
  author    = {{Wolfram Research}},
  title     = {Mathematica},
  version   = {14.3},
  year      = {2026},
  location  = {Champaign, IL},
  publisher = {Wolfram Research, Inc.}
}

@software{Python,
author = {{Python Software Foundation}},
title = {Python Language Reference, version 3.13.9},
year = {2025},
url = {https://www.python.org}
}

@article{NumPy,
author = {Harris, Charles R. and Millman, K. Jarrod and van der Walt, St{'e}fan J. and others},
title = {Array programming with NumPy},
journal = {Nature},
volume = {585},
pages = {357--362},
year = {2020},
doi = {10.1038/s41586-020-2649-2}
}

@article{SciPy,
  author  = {Virtanen, Pauli and Gommers, Ralf and Oliphant, Travis E. and Haberland, Matt and others},
  title   = {SciPy 1.0: fundamental algorithms for scientific computing in Python},
  journal = {Nat. Methods},
  volume  = {17},
  pages   = {261--272},
  year    = {2020},
  doi     = {10.1038/s41592-019-0686-2}
}

@article{kellerModelFrameworkIon2025,
  title = {A {{model framework}} for {{ion channels}} with {{selectivity filters based}} on {{non-equilibrium thermodynamics}}},
  author = {Keller, Christine and Landstorfer, Manuel and Fuhrmann, Jürgen and Wagner, Barbara},
  date = {2025-09-20},
  journaltitle = {Entropy},
  shortjournal = {Entropy},
  volume = {27},
  number = {9},
  pages = {981},
  issn = {1099-4300},
  doi = {10.3390/e27090981},
  url = {https://www.mdpi.com/1099-4300/27/9/981},
  urldate = {2026-03-18},
  langid = {english}
}

@article{zhengSecondorderPoissonNernst2011a,
  title = {Second-order {{Poisson}}–{{Nernst}}–{{Planck}} solver for ion transport},
  author = {Zheng, Qiong and Chen, Duan and Wei, Guo-Wei},
  date = {2011-06},
  journaltitle = {J. Comput. Phys.},
  shortjournal = {Journal of Computational Physics},
  volume = {230},
  number = {13},
  pages = {5239--5262},
  issn = {00219991},
  doi = {10.1016/j.jcp.2011.03.020},
  url = {https://linkinghub.elsevier.com/retrieve/pii/S002199911100163X},
  urldate = {2026-03-18},
  langid = {english}
}

@article{singerSingularPerturbationAnalysis2008,
  title = {Singular perturbation analysis of the steady-state {{Poisson}}–{{Nernst}}–{{Planck}} system: {{applications}} to ion channels},
  shorttitle = {Singular Perturbation Analysis of the Steady-State {{Poisson}}–{{Nernst}}–{{Planck}} System},
  author = {Singer, A. and Gillespie, D. and Norbury, J. and Eisenberg, R. S.},
  date = {2008-10},
  journaltitle = {Eur. J. Appl. Math.},
  shortjournal = {Eur. J. Appl. Math},
  volume = {19},
  number = {5},
  pages = {541--560},
  issn = {0956-7925, 1469-4425},
  doi = {10.1017/S0956792508007596},
  url = {https://www.cambridge.org/core/product/identifier/S0956792508007596/type/journal_article},
  urldate = {2025-06-02},
  langid = {english}
}

@article{messiasAssessingImpactValence2022,
  title = {Assessing the impact of valence asymmetry in ionic solutions and its consequences on the performance of supercapacitors},
  author = {Messias, Andresa and Fileti, Eudes E.},
  date = {2022},
  journaltitle = {Phys. Chem. Chem. Phys.},
  shortjournal = {Phys. Chem. Chem. Phys.},
  volume = {24},
  number = {34},
  pages = {20445--20453},
  issn = {1463-9076, 1463-9084},
  doi = {10.1039/D2CP00348A},
  url = {https://xlink.rsc.org/?DOI=D2CP00348A},
  urldate = {2026-02-05},
  langid = {english}
}

@book{benderAdvancedMathematicalMethods2009,
  title = {Advanced Mathematical Methods for Scientists and Engineers. 1: {{Asymptotic}} Methods and Perturbation Theory},
  shorttitle = {Advanced Mathematical Methods for Scientists and Engineers. 1},
  author = {Bender, Carl M. and Orszag, Steven A.},
  date = {2009},
  edition = {Nachdr.},
  publisher = {Springer},
  location = {New York, NY},
  isbn = {978-0-387-98931-0},
  langid = {english},
  pagetotal = {593}
}

@book{hibbertIntroductionElectrochemistry1993,
  title = {Introduction to Electrochemistry},
  author = {Hibbert, D. Brynn},
  date = {1993},
  series = {{{MacMillan}} Physical Science Series},
  edition = {1. publ},
  publisher = {Macmillan},
  location = {Basingstoke},
  isbn = {978-0-333-56303-8},
  langid = {english},
  pagetotal = {350}
}

@article{kovalenkoTheoreticalInvestigationPhenomenon2022,
  title = {Theoretical {{investigation}} of the {{phenomenon}} of {{space charge breakdown}} in {{electromembrane systems}}},
  author = {Kovalenko, Anna and Chubyr, Natalia and Uzdenova, Aminat and Urtenov, Makhamet},
  date = {2022-10-26},
  journaltitle = {Membranes},
  shortjournal = {Membranes},
  volume = {12},
  number = {11},
  pages = {1047},
  issn = {2077-0375},
  doi = {10.3390/membranes12111047},
  url = {https://www.mdpi.com/2077-0375/12/11/1047},
  urldate = {2026-04-08},
  langid = {english}
}

@article{mendheReviewElectrolytesSupercapacitor2023,
  title = {A review on electrolytes for supercapacitor device},
  author = {Mendhe, Arpit and Panda, H. S.},
  date = {2023-10-26},
  journaltitle = {Discov. Mater.},
  shortjournal = {Discov Mater},
  volume = {3},
  number = {1},
  pages = {29},
  issn = {2730-7727},
  doi = {10.1007/s43939-023-00065-3},
  url = {https://link.springer.com/10.1007/s43939-023-00065-3},
  urldate = {2026-04-08},
  langid = {english}
}

@article{guptaDiffusiophoreticDiffusioosmoticVelocities2019,
  title = {Diffusiophoretic and diffusioosmotic velocities for mixtures of valence-asymmetric electrolytes},
  author = {Gupta, Ankur and Rallabandi, Bhargav and Stone, Howard A.},
  date = {2019-04-12},
  journaltitle = {Phys. Rev. Fluids},
  shortjournal = {Phys. Rev. Fluids},
  volume = {4},
  number = {4},
  pages = {043702},
  issn = {2469-990X},
  doi = {10.1103/PhysRevFluids.4.043702},
  url = {https://link.aps.org/doi/10.1103/PhysRevFluids.4.043702},
  urldate = {2026-02-05},
  langid = {english}
}

@article{messiasSaltinwaterWaterinsaltElectrolytes2022,
  title = {Salt-in-water and water-in-salt electrolytes: the effects of the asymmetry in cation and anion valence on their properties},
  shorttitle = {Salt-in-Water and Water-in-Salt Electrolytes},
  author = {Messias, Andresa and C. Da Silva, Débora A. and Fileti, Eudes E.},
  date = {2022},
  journaltitle = {Phys. Chem. Chem. Phys.},
  shortjournal = {Phys. Chem. Chem. Phys.},
  volume = {24},
  number = {1},
  pages = {336--346},
  issn = {1463-9076, 1463-9084},
  doi = {10.1039/D1CP04259A},
  url = {https://xlink.rsc.org/?DOI=D1CP04259A},
  urldate = {2026-02-05},
  langid = {english}
}

@article{henriqueImpactAsymmetriesValences2022a,
  title = {Impact of asymmetries in valences and diffusivities on the transport of a binary electrolyte in a charged cylindrical pore},
  author = {Henrique, Filipe and Zuk, Pawel J. and Gupta, Ankur},
  date = {2022-11},
  journaltitle = {Electrochim. Acta},
  shortjournal = {Electrochimica Acta},
  volume = {433},
  pages = {141220},
  issn = {00134686},
  doi = {10.1016/j.electacta.2022.141220},
  url = {https://linkinghub.elsevier.com/retrieve/pii/S0013468622013779},
  urldate = {2026-04-08},
  langid = {english}
}

@article{baluElectrochemicalImpedanceSpectrum2022,
  title = {The electrochemical impedance spectrum of asymmetric electrolytes across low to moderate frequencies},
  author = {Balu, Bhavya and Khair, Aditya S.},
  date = {2022-04},
  journaltitle = {J. Electroanal. Chem.},
  shortjournal = {Journal of Electroanalytical Chemistry},
  volume = {911},
  pages = {116222},
  issn = {15726657},
  doi = {10.1016/j.jelechem.2022.116222},
  url = {https://linkinghub.elsevier.com/retrieve/pii/S1572665722002144},
  urldate = {2026-04-08},
  langid = {english}
}

@article{vansoestbergenDiffusechargeEffectsTransient2010,
  title = {Diffuse-charge effects on the transient response of electrochemical cells},
  author = {van Soestbergen, M. and Biesheuvel, P. M. and Bazant, M. Z.},
  journal = {Phys. Rev. E},
  volume = {81},
  issue = {2},
  pages = {021503},
  numpages = {13},
  year = {2010},
  month = {Feb},
  publisher = {American Physical Society},
  doi = {10.1103/PhysRevE.81.021503},
  url = {https://link.aps.org/doi/10.1103/PhysRevE.81.021503}
}

@article{Nernst+1904+52+55,
  title = {Theorie der reaktionsgeschwindigkeit in heterogenen systemen},
  author = {Nernst, W.},
  date = {1904},
  journaltitle = {Z. Phys. Chem.},
  volume = {47U},
  number = {1},
  pages = {52--55},
  doi = {doi:10.1515/zpch-1904-4704},
  url = {https://doi.org/10.1515/zpch-1904-4704},
  urldate = {2026-03-18}
}

@article{chaoIntegralEquationMethod2023,
  title = {Integral equation method for the {{1D}} steady-state {{Poisson--Nernst--Planck}} equations},
  author = {Chao, Zhen and Geng, Weihua and Krasny, Robert},
  date = {2023-10},
  journaltitle = {J. Comput. Electron.},
  shortjournal = {J Comput Electron},
  volume = {22},
  number = {5},
  pages = {1396--1408},
  issn = {1569-8025, 1572-8137},
  doi = {10.1007/s10825-023-02092-y},
  url = {https://link.springer.com/10.1007/s10825-023-02092-y},
  urldate = {2025-05-19},
  langid = {english}
}

@software{georgina_ryan_2026_19695204,
  author       = {Georgina Ryan},
  title        = {georginaryan/asymmetric-valence-electrolytes},
  month        = apr,
  year         = 2026,
  publisher    = {Zenodo},
  version      = {v.1.0.0},
  doi          = {10.5281/zenodo.19695204},
  url          = {https://doi.org/10.5281/zenodo.19695204}
}

@incollection{raabIntegrationFiniteTerms2022,
author    = {Khovanskii, Askold},
  title     = {Comments on J. F. Ritt's Book \textit{Integration in Finite Terms}},
  booktitle = {Integration in Finite Terms: Fundamental Sources},
  editor    = {Raab, C. G. and Singer, M. F.},
  pages     = {137--202},
  publisher = {Springer},
  address   = {Cham},
  year      = {2022}
}

\end{document}

\maketitle

\section*{Dirichlet concentration boundary conditions}\label{SuppMaterialCBC}
Dirichlet concentration boundary conditions are a common choice for modelling ion channels in biological cells with the Poisson--Nernst--Planck equations\autocite{kellerModelFrameworkIon2025,zhengSecondorderPoissonNernst2011a,singerSingularPerturbationAnalysis2008,wangSingularPerturbationSolutions2014,chenQualitativePropertiesSteadyState1997}. 
Here we generalise the approach taken by  \citet{chenQualitativePropertiesSteadyState1997} for a $1\R1$ electrolyte with Dirichlet concentration boundary conditions to produce analytic expressions for the composite solutions in the $\rr=2$ and $\rr=1/2$ electrolyte cases. In this derivation, we refer back to elements of the main text that are duplicated in this problem and explicitly present details of the different aspects.\\

We solve the following dimensionless problem after applying the scalings defined in [6],
\begin{subequations}\label{ConcGovEqns}
\begin{align}
\frac{\mathrm{d} \pOG}{\mathrm{d} x}+ \rr\pOG\frac{\mathrm{d} \potOG}{\mathrm{d} x}&=-I_p\label{NEqn2},\\
\frac{\mathrm{d} \nOG}{\mathrm{d} x}- \nOG\frac{\mathrm{d} \potOG}{\mathrm{d} x}&=I_n\label{PEqn2}, \\
\eps^2\frac{\mathrm{d}^2\potOG}{\mathrm{d} x^2} &=-\pOG+\nOG, \label{Poisson2}
\end{align}
\end{subequations}
with the dimensionless boundary conditions,
\begin{align}
\label{ConcBCs}
    p(0)=p_L, \quad p(1)=p_R, \quad n(0)=n_L, \quad n(1)=n_R,\quad \potOG(0)= V,  \quad \potOG(1)=0,
    \refstepcounter{equation}
\tag{\theequation a-f}
\end{align}
where $p_L,p_R,n_L,n_R>0$. In contrast to the problem considered in the main text, the ionic fluxes $I_p,I_n$ in the Nernst--Planck equations (\ref{ConcGovEqns}a,b)  are unknown constants. 

\subsection*{Electroneutral Bulk}
We no longer apply a mass-conservation condition upon integrating [19], so we find 
\begin{equation}
\C(x)=-2\B x+\cone. \label{Cout2FluxApp}
\end{equation}
for some constant $\cone>0$. 
Now, solving [21], we find
\begin{equation}
\pot(x)=\frac{\II}{2\B } \log\left( -2\B x+\cone \right)+\ctwo,\label{PhiOut1AsymApp}
\end{equation}
where we have introduced a new integration constant $\ctwo$.

\subsection*{Left Diffuse Layer}
This problem has the following matching and boundary conditions:
\begin{subequations}\label{MatchingBCsLeftApp}
\begin{align}
\potL\left(\infty\right)&=\phi_0(0)=\frac{\II}{2\B}\log\left( \cone\right)+\ctwo ,\label{Phi0ConditionL2App}\\
\pL(\infty)&=\C(0)=\cone ,\label{p0ConditionL2App}\\
\nL(\infty)&=\C(0)=\cone,\label{n0ConditionL2App}\\
\potL(0)&=V,\\
\pL(0)&=p_L,\\
\nL(0)&=n_L,
\end{align}
\end{subequations}
where $\cone$ is to be determined.  Again, we have Boltzmann relations at leading order given by [27].

Applying the matching conditions (\ref{MatchingBCsLeftApp}a--c), we find
\begin{subequations}
    \begin{align}
        p_L=\frac{\cone^2}{n_L}\exp[(\rr-1)(\pot(0)-V)],\label{pLDef2BApp}\\
        \pot(0)=V+\frac{1}{1+\rr}\log\left( \frac{p_{L}}{n_{L}} \right). \label{potL0App}
    \end{align}
\end{subequations}

Therefore, substituting \eqref{potL0App} into \eqref{pLDef2BApp}, we find
\begin{equation} \label{CBC constant def}
    \cone=p_L^{1/(1+\rr)} n_L^{r/(1+\rr)}.
\end{equation}
Note that this equation determines the degree of freedom $\cone$ in terms of the given boundary conditions $p_L, n_L$. Hence, we can fix $\ctwo$ by equating \eqref{Phi0ConditionL2App} and \eqref{potL0App} and substituting in $\cone$ from \eqref{CBC constant def},
\begin{equation}
    \ctwo=V+ \frac{1}{1+\rr}\log\left(\frac{p_L}{n_L}\right)-\frac{\II}{2\B(1+\rr)} \log(p_L n_L^r).
\end{equation}
Therefore, the outer layer potential is given by 
\begin{equation}
   \pot(x)=V+\frac{\II}{2\B} \log\left( -2\B x+ p_L^{1/(1+\rr)} n_L^{r/(1+\rr)}\right)-\frac{\II}{2\B(1+\rr)}\log\left( p_L n_L^{r} \right) + \frac{1}{1+\rr}\log \left(\frac{p_L}{n_L}\right), \label{fullOuterPotential2CBC}
\end{equation}
Now, we consider the Poisson equation [23c],
\begin{equation}
\dd{^2\potL}{\xL^2}=-p_L\exp(\rr(V-\potL(\xL)))+n_L\exp(\potL(\xL)-V). \label{SubstitutedLeftEqnApp}
\end{equation}
We then multiply \eqref{SubstitutedLeftEqnApp} by $\mathrm{d}\potL/\mathrm{d}\xL$ and integrate. The integration constant is fixed by the matching conditions \eqref{MatchingBCsLeftApp}, the definition of $\phi_{0}(0)$ \eqref{potL0App}, and the implication of the matching conditions that ${\potL'(\infty )=0}$.
Hence, we find that 
\begin{align}
\left(\frac{\mathrm{d}\potL}{\mathrm{d}\xL}\right)^2= 2n_L \left( \frac{p_L}{r n_L }\exp(\rr(V-\potL(\xL)))+\exp(\potL(\xL)-V)-\frac{(1+\rr)}{\rr} \left(\frac{p_L}{n_L}\right)^{1/(1+r)}\right).  
\label{diffeqnApp}
\end{align}

It is helpful to convert the nonlinear functions of $\potL$ in \eqref{diffeqnApp} into powers of a transformed variable. Therefore, we make the substitution \begin{equation} \label{mSubApp}
     \mL(\xL)=\exp((\potL(\xL)-V)/2).
\end{equation}  Under this substitution, the relevant boundary and matching conditions from \eqref{MatchingBCsLeftApp} become
\begin{equation} \label{m BCs}
    \mL(0)=1, \quad \mL(\infty)=\exp((\potOG(0)-V)/2)=\tl,\refstepcounter{equation}\tag{\theequation \text{a,b}}
\end{equation} where
\begin{equation}
    \tl=\left(\frac{p_L}{n_L}\right)^{1/(2(1+\rr))}>0.\label{alpha Asym def CBC}
\end{equation} 
 With this substitution \eqref{mSubApp} and taking the square root, \eqref{diffeqnApp} becomes
\begin{align}
\frac{\mathrm{d}\mL}{\mathrm{d}\xL}= \text{sign}(\tl-1)\sqrt{\frac{n_L}{2}}\sqrt{\frac{\tl^{2(1+\rr)}}{r}\mL^{2-2 r}+ \mL^{4}-\frac{1+\rr}{\rr} \tl^2 \mL^2}. \label{Gen LBL eqnApp}
\end{align}
\\
This equation \eqref{Gen LBL eqnApp} is the same equation we solved in the mass-conservation case [36] with the substitution of $\tl$ for $\alpha_L$. We find the same parallel in the right boundary layer. Therefore, the boundary layer solutions for this problem are simply the main-text solutions from the $\rr=1/2$ [59] and $\rr=2$ [56] cases as well as the general asymmetric valence solutions [37] and [44] under the substitution of ($\tl,\tr)$ for $(\alpha_L,\alpha_R)$, where
\begin{equation}
    \tr=\left(\frac{p_R}{n_R}\right)^{1/(2(1+\rr))}>0.\label{alpha Asym def CBC right}
\end{equation} 

\subsection*{Right Diffuse Layer}
To close the problem, it remains to use information from the right boundary layer to determine the constants $I_p,I_n$ in terms of the known boundary concentrations, and hence to define $\II$ and $\B$. \\

Applying the equivalent matching conditions in the right boundary layer, we find 
\begin{align}
\potR(\infty)&=\phi_0(1)=V+\frac{\II}{2\B} \log\left( -2\B + (p_L n_L^{r})^{1/(1+\rr)}\right)-\frac{\II}{2\B(1+\rr)}\log\left( p_L n_L^{r} \right) + \frac{1}{1+\rr}\log \left(\frac{p_L}{n_L}\right), \label{Matching Right Inf}\\
    \pR(\infty)&=\nR(\infty)=p_L^{1/(1+\rr)} n_L^{r/(1+\rr)}-2\B.
\end{align}
By multiplying and dividing the Boltzmann relations for the concentration in the right boundary layer (an equivalent procedure to the main text), we find that
\begin{align}
\refstepcounter{equation}%
\global\edef\RightPotentialCBCNumber{\arabic{equation}}% store numeric parent
\phantomsection\label{RightPotentialCBCAnchor}
    \potR(\infty)=\pot(1)=\frac{1}{1+\rr}\log\left( \frac{p_{R}}{n_{R}} \right), \quad \quad p_R=\frac{(p_L^{1/(1+\rr)} n_L^{r/(1+\rr)}-2\B)^{(1+\rr)}}{n_R^{\rr}}.\tag{\RightPotentialCBCNumber a,b}
\end{align}

Rearranging \hyperref[RightPotentialCBCAnchor]{(\RightPotentialCBCNumber b)}, we find that 
\begin{equation}
    \B=\frac{1}{2}\left(p_L^{1/(1+\rr)} n_L^{\rr/(1+\rr)}- p_R^{1/(1+\rr)} n_R^{\rr/(1+\rr)}\right). \label{J definition}
\end{equation}

Therefore, equating the definitions of $\pot(1)$ from \eqref{Matching Right Inf} and \eqref{fullOuterPotential2CBC}, we find that 
\begin{comment}
=\frac{\LL}{\log(1-\frac{2\B}{\cone})} \left[\frac{1}{\K}\log\left(\frac{p_R}{n_R}\right)-V-\frac{1}{z_n}\log\left(\frac{\cone}{n_L}\right)\right]\\
\end{comment}
\begin{align}
    \II&
    = \frac{2\B \left( \log\left(\frac{n_Lp_R}{n_Rp_L}\right)-(1+\rr) V\right)}{\rr\log(n_R/n_L)+\log(p_R/p_L)}. \label{II definition CBC}
\end{align}
This matches equation (44) for \mbox{$1\R1$} electrolytes in \citet{chenQualitativePropertiesSteadyState1997} (noting that in their notation, ${I^{(0)}=2\II|_{\rr=1}}$ and ${J^{(0)}=-4\B|_{\rr=1}}$). We can therefore fully close the problem, expressing $I_p$ and $I_n$ in terms of $\B$ \eqref{J definition} and $\II$ \eqref{II definition CBC},
\begin{align}
\refstepcounter{equation}%
\global\edef\IPINnumber{\arabic{equation}}% store numeric parent
\phantomsection\label{IPINAnchor}
    I_p=r\II+2\B, \quad I_n=\II -2\B.
    \tag{\IPINnumber a,b}
\end{align}

\section*{Asymptotic composite solutions}
We provide the elements of the asymptotic composite solutions for the $\rr=2$ and $\rr=1/2$. For completeness, we additionally provide the solution for the $\rr=1$ case (matching the \mbox{$1\R 1$} solution from \citet{chenQualitativePropertiesSteadyState1997}) and the general electrolyte solution. 
\subsection*{\texorpdfstring{$\rr=2$ solution}{r=2 solution}}
The outer solution for the potential of $\rr=2$ electrolytes is 
\begin{equation}
    \pot(x)=V+\frac{1}{3} \log \left(\frac{p_L}{n_L}\right)-\frac{\log \left(\frac{n_R p_L}{n_L p_R}\right)+3 V }{2 \log \left(\frac{n_R}{n_L}\right)+\log \left(\frac{p_R}{p_L}\right)}\log \left(x \left[{\left(\frac{n_R^2 p_R}{n_L^2 p_L}\right)^{1/3}}-1\right]+1\right),
\end{equation}
and the outer solution for the concentrations is 
\begin{equation}
    c_0(x)=n_L^{2/3} p_L^{1/3}-x \left(n_L^{2/3} p_L^{1/3}-n_R^{2/3} p_R^{1/3}\right).
\end{equation}
The boundary layer solutions for the potential are 
\begin{subequations}
\begin{align}
    \Phi(\xL)=V+\log \left(\frac{\tl^2}{2}\right)+\log\left( 3 \left(\frac{\tanh\left( \dfrac{\tl\sqrt{3n_L}}{2}  \xL\right)+\left(\sqrt{\dfrac{2}{3 \tl ^2}+\dfrac{1}{3}}\right)}{1+\left(\sqrt{\dfrac{2}{3 \tl ^2}+\dfrac{1}{3}}\right)\tanh\left( \dfrac{\tl\sqrt{3 n_L}}{2}  \xL\right)}\right)^2-1\right).\\
    \varphi(\xR)=\log\left(\frac{\tr^2}{2}\right)+\log\left( 3 \left(\frac{\tanh\left( \dfrac{\tr\sqrt{3n_R}}{2}  \xR\right)+\left(\sqrt{\dfrac{2}{3 \tr ^2}+\dfrac{1}{3}}\right)}{1+\left(\sqrt{\dfrac{2}{3 \tr ^2}+\dfrac{1}{3}}\right)\tanh\left( \dfrac{\tr\sqrt{3n_R}}{2}  \xR\right)}\right)^2-1\right),
    \end{align}
\end{subequations}
where
\begin{equation}
    \tl=\left(\frac{p_L}{n_L}\right)^{1/6}, \quad \tr=\left(\frac{p_R}{n_R}\right)^{1/6}.
\end{equation}

\subsection*{\texorpdfstring{$\rr=1/2$ solution}{r=1/2 solution}}
The outer solution for the potential of $\rr=1/2$ electrolytes is
\begin{equation}
    \pot(x)=V+\frac{2}{3} \log \left(\frac{p_L}{n_L}\right)-\frac{2\log \left(\frac{n_R p_L}{n_L p_R}\right)+3 V }{\log \left(\frac{n_R}{n_L}\right)+2\log \left(\frac{p_R}{p_L}\right)}\log \left(x \left[{\left(\frac{n_R p_R^2}{n_L p_L^2}\right)^{1/3}}-1\right]+1\right),
\end{equation}
and the outer solution for the concentrations is
\begin{equation}
    c_0(x)=n_L^{1/3} p_L^{2/3}-x \left(n_L^{1/3} p_L^{2/3}-n_R^{1/3} p_R^{2/3}\right).
\end{equation}
The boundary layer solutions for the potential are
\begin{subequations}
\begin{align}
  \Phi(\xL)=V+2\log\left(  \frac{
2\tl \left( \tanh\!\left(\frac{\sqrt{6 n_L}}{4}\tl \xL \right)
+ \sqrt{\frac{3}{1+2 \tl}}\right)^{2}
}{
3\left(1 + \sqrt{\frac{3}{1+2\tl}}
\,\tanh\!\left(\frac{\sqrt{6n}}{4}\, \tl\xL \right) \right)^2-
\left(\tanh\!\left(\frac{\sqrt{6n_L}}{4}\, \tl\xL\right)
+ \sqrt{\frac{3}{1+2\tl}}\right)^2
}\right),\\
\label{12rightb}
    \varphi(\xR)=2\log\left(  \frac{
2\tr \left( \tanh\!\left(\frac{\sqrt{6 n_R}}{4}\tr \xR \right)
+ \sqrt{\frac{3}{1+2 \tr}}\right)^{2}
}{
3\left(1 + \sqrt{\frac{3}{1+2\tr}}
\,\tanh\!\left(\frac{\sqrt{6 n_R}}{4}\,  \tr \xR\right) \right)^2-
\left(\tanh\!\left(\frac{\sqrt{6n_L}}{4}\,  \tr \xR\right)
+ \sqrt{\frac{3}{1+2\tr}}\right)^2
}\right),
\end{align}
\end{subequations}
where
\begin{equation}
    \tl=\left(\frac{p_L}{n_L}\right)^{1/3}, \quad \tr=\left(\frac{p_R}{n_R}\right)^{1/3}.
\end{equation}

\subsection*{\texorpdfstring{$\rr=1$ solution}{r=1 solution}}
For completeness, we provide the $\rr=1$ result. The outer solution for the potential is given by
\begin{equation}
    \pot(x)=V+\frac{1}{2} \log \left(\frac{p_L}{n_L}\right)-\frac{\log \left(\frac{n_R p_L}{n_L p_R}\right)+2 V }{\log \left(\frac{n_R}{n_L}\right)+\log \left(\frac{p_R}{p_L}\right)}\log \left(x \left[{\left(\frac{n_R p_R}{n_L p_L}\right)^{1/2}}-1\right]+1\right),
\end{equation}
and the outer solution for the concentrations is
\begin{equation}
    c_0(x)=n_L^{1/2} p_L^{1/2}-x \left(n_L^{1/2} p_L^{1/2}-n_R^{1/2} p_R^{1/2}\right).
\end{equation}
The solutions for the potential in the boundary layers are 
\begin{subequations}
    \begin{align} 
             \potL(\xL)&=V+2\log\left( \frac{\tl\left(1+\tl \tanh\left(\tl \sqrt{\dfrac{  n_L}{2}}\xL\right)\right)}{\tl+ \tanh\left(\tl  \sqrt{\dfrac{n_L}{2}}\xL\right)}\right), \label{r1 left}\\
             \potR(\xR)&=2\log\left( \frac{\tr\left(1+\tr \tanh\left(\tr  \sqrt{\dfrac{ n_R }{2}}\xR\right)\right)}{\tr+ \tanh\left(\tr  \sqrt{\dfrac{ n_R}{2}}\xR\right)}\right),
            \label{r1 right}
    \end{align}
\end{subequations}
where
\begin{equation}
    \tl=\left(\frac{p_L}{n_L}\right)^{1/4}, \quad \tr=\left(\frac{p_R}{n_R}\right)^{1/4}.
\end{equation}

\subsection*{General asymmetric binary electrolyte}
The outer solution for the potential is given by
\begin{equation}
    \pot(x)=V+\frac{1}{1+\rr} \log \left(\frac{p_L}{n_L}\right)-\frac{\log \left(\frac{n_R p_L}{n_L p_R}\right)+(1+\rr) V }{\rr\log \left(\frac{n_R}{n_L}\right)+\log \left(\frac{p_R}{p_L}\right)}\log \left(x \left[{\left(\frac{n_R^r p_R}{n_L^r p_L}\right)^{1/(1+\rr)}}-1\right]+1\right),
\end{equation}
and the outer solution for the concentrations is
\begin{equation}
    c_0(x)=(n_L^{r} p_L)^{1/(1+r)}-x \left(\left(n_L^rp_L \right)^{1/(1+r)}-\left(n_R^rp_R \right)^{1/(1+r)}\right).
\end{equation}
The solutions for the potential in the boundary layers are 
\begin{subequations}
    \begin{align} 
    \xL
&= \text{sign}(\alpha_L-1)\,
\sqrt{\frac{2}{n_L}}
\int_{1}^{\mL(\xL)}
\left(
\frac{\tl^{2(1+\rr)}}{\rr} s^{\,2-2 \rr}
+ s^{4 }
-\frac{1+\rr}{\rr} \tl^2 s^{\,2}
\right)^{-1/2}
\,\mathrm{d}s,\\
          \xR&
= \text{sign}(\alpha_R-1)\,
{\sqrt\frac{2}{n_R}}
\int_{1}^{\mR(\xR)}
\left(
\frac{\tr^{2(1+\rr)}}{\rr} s^{\,2-2\rr}
+s^{4}
-\frac{(1+\rr)}{\rr} \tr^2 s^{\,2}
\right)^{-1/2}
\,\mathrm{d}s .
            \label{gen APP R}
    \end{align}
\end{subequations}
where
\begin{equation}
    \tl=\left(\frac{p_L}{n_L}\right)^{1/4}, \quad \tr=\left(\frac{p_R}{n_R}\right)^{1/4}.
\end{equation}
Note that $\Phi(\xL)$ is related to  $\mL(\xL)$ by \eqref{mSubApp} and $\mR(\xR)=\exp(\potR(\xR)/2).$
\sloppy

\printbibliography